\documentclass[epj]{svjour}
\usepackage{amsmath,amssymb}
\usepackage[hiresbb,final]{graphicx}
\usepackage{hyphenat} 

\newcommand{\PDMS}{\text{PD}}
\newcommand{\air}{\text{air}}
\newcommand{\eq}{\text{eq}}
\newcommand{\td}{\text{td}}
\newcommand{\PEW}{\text{PEW}}
\newcommand{\RSS}{\text{RSS}}

\newcommand{\lef}{{(\ell)}}
\newcommand{\rig}{{(r)}}
\newcommand{\vol}{\nu}
\newcommand{\Pe}{\mbox{\sffamily\textit{Pe}}}
\newcommand{\Peloc}{\mbox{\sffamily\textit{Pe}}_\textit{loc}}
\newcommand{\PhCoeff}{\mathcal{L}}
\newcommand{\chpot}{\gamma}
\newcommand{\bfe}{\mathbf{e}} 
\newcommand{\bfg}{\mathbf{g}}
\newcommand{\bfJ}{\mathbf{J}} 
\newcommand{\bfj}{\mathbf{j}} 
\newcommand{\bfN}{\mathbf{N}} 
\newcommand{\bfr}{\mathbf{r}} 
\newcommand{\bfU}{\mathbf{U}} 
\newcommand{\bfv}{\mathbf{v}} 
\newcommand{\bfnabla}{\boldsymbol{\nabla}}
\newcommand{\barc}{\bar{c}} 
\newcommand{\barL}{\bar{\PhCoeff}} 
\newcommand{\barD}{\bar{D}} 
\newcommand{\barchpot}{\bar{\chpot}} 
\newcommand{\barbfJ}{\bar{\mathbf{J}}} 
\newcommand{\barbfj}{\bar{\mathbf{j}}} 
\newcommand{\vavg}{{0}} 
\newcommand{\dotp}{\boldsymbol{\cdot}}

\newcommand{\grad}{\bfnabla}
\renewcommand{\div}{\operatorname{div}}

\newcommand{\micrometer}{\mu\textrm{m}}
\newcommand{\centimeter}{\textrm{cm}}
\allowdisplaybreaks
\begin{document}

\title{Modeling phase behavior for quantifying micro-pervaporation experiments}
\author{Michael Schindler\and Armand Ajdari}
\institute{Laboratoire PCT, UMR~``Gulliver'' CNRS-ESPCI 7083, 10 rue Vauquelin, 75231 Paris cedex 05}
\date{Received: date / Revised version: date}

\abstract{%
We present a theoretical model for the evolution of mixture concentrations in a
micro-pervaporation device, similar to those recently presented experimentally.
The described device makes use of the pervaporation of water through a thin PDMS
membrane to build up a solute concentration profile inside a long microfluidic
channel. We simplify the evolution of this profile in binary mixtures to a
one-dimensional model which comprises two concentration-dependent coefficients.
The model then provides a link between directly accessible experimental
observations, such as the widths of dense phases or their growth velocity, and
the underlying chemical potentials and phenomenological coefficients. It shall
thus be useful for quantifying the thermodynamic and dynamic properties of
dilute and dense binary mixtures.
}

\maketitle

\section{Introduction}
The thermodynamic and kinetic behavior of dense solutions is of central interest
in a number of industrial and scientific activities. The so\hyp{}called
``formulation'' of multi\hyp{}component systems for the production of cosmetics,
paints and also comestible goods are industrial examples~\cite{industry}.
Scientific interests range from mobilities of macromolecules in cells to the
statistical physics and material properties of colloidal
suspensions~\cite{cells,RusSavSch89} and other mixtures. In all these
applications it is important to know the diffusivity or mobility of solutes,
their permeability with respect to a solvent or other solutes and similar
non\hyp{}equilibrium quantities. Eventually, dense mixtures are complex substances
which are likely to undergo a number of thermodynamic and dynamic transitions of
state, which makes their analysis a challenge.
\begin{figure}[b!]
  \centering
  \includegraphics{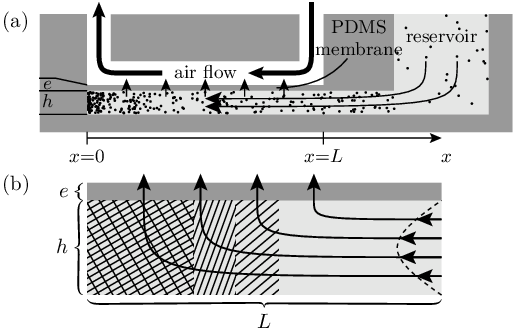}%
  \caption{(a)~Schematic side view of the micro\hyp{}evaporator device. It comprises
  a thin PDMS~membrane (thickness $e{\approx}20\micrometer$) which allows water
  to pervaporate out of the lower channel which is filled by a solution
  ($h{\approx}100\micrometer$, $L{\approx}1\centimeter$). (b)~More details of
  the flow pattern: The pervaporating solvent induces also a concentration
  profile of trapped solute at the end of the channel, here in several dense
  phases as indicated by the different patterns.}%
  \label{fig:device}%
\end{figure}%
%

The thermodynamic and dynamic properties of dilute and dense mixtures are
usually measured by experimental techniques such as
sedimentation/centrifugation, ultra\hyp{}filtration or reverse
osmosis~\cite{PepEllWor05,LebLewSch02,Paul04}. Depending on the properties of the solution
under investigation, one or the other technique proves to be advantageous.
Recently, a promising new microfluidic tool for the analysis of aqueous
solutions has been introduced, the
\emph{microevaporator}~\cite{LenLonTabJoaAjd06,LenJoaAjd07}. The employed method
is applicable to a wide range of different solutes, such as electrolytes,
surfactants, colloids, or polymers. It makes use of the pervaporation of the
solvent water through a thin membrane (short vertical arrows in
Fig.~\ref{fig:device}a). The solutes cannot pass the membrane and build up a
concentration profile in an underlying microfluidic channel (black dots in
Fig.~\ref{fig:device}a). This profile of solute concentration contains
information on the transport coefficients of the mixture. In experiments, the
method has been successfully employed for the semi\hyp{}quantitative screening of
equilibrium phase diagrams and the determination of transport coefficients near
equilibrium~\cite{SalLen08}. As a tool for determining phase diagrams, the
microevaporator resembles the functionality of the ``phase
chip''~\cite{Fraden,Baitz}. However, the microevaporator allows to go one step
further towards the controlled analysis of out\hyp{}of\hyp{}equilibrium situations.

The quantitative interpretation of the experimental observations in the
microevaporator, that is the extraction of the transport coefficients, requires
a theoretical model. The observed quantities are spatio\hyp{}temporal profiles of
solute densities, the texture of occurring phases, and the growth dynamics of
dense phases which present a moving phase interface. The link between these
observables on one side and the underlying thermodynamic equation of state, the
transport coefficients of the mixture, and possibly more detailed kinetic
parameters on the other side can only be achieved by a sufficiently detailed
theoretical model. At the moment, no theoretical model exists for the
concentration process caused by pervaporation. We here start to fill this gap.

A model for the phase behavior in microevaporation is subject to the following
constraints: On one hand, the microevaporator device can be applied to very
different types of solutes, such as colloids, surfactants and electrolytes.
Consequently, a sufficiently \emph{general} theoretical modeling is required
which includes phase transitions and out\hyp{}of\hyp{}equilibrium phenomena
namely metastable phases and precursor phases. It must further take into account
the influences of the driving conditions and of the design parameters of the
channels. On the other hand, the model should turn the microevaporator into a
\emph{quantitative} tool, making the extraction of thermodynamic and kinetic
properties of mixtures possible in a systematic and unambiguous way. This aim
requires a sufficiently simple model which permits to approach several limits
analytically, and which lends itself to a fast numerical solution of the
equations.

\emph{The model proposed here can thus be seen as a first compromise between the
complexity of the subject, including out\hyp{}of\hyp{}equilibrium properties, and the
requirement to allow a fast and direct comparison of its solutions with
experimental observations.}

In order not to overload the description, we restrict it to binary mixtures
(water plus a single solute) which are described by two spatio\hyp{}temporal fields,
namely the concentration of solute~$\phi(x,t)$ and a velocity $v(x,t)$ of the
mixture. The evolution of these two variables is governed by equations of the
convection--diffusion type, which comprise two transport coefficients, namely
$D(\phi)$~for inter\hyp{}diffusion of solute and water, and $q(\phi)$ for the
pervaporation. Both coefficients are state\hyp{}dependent, which is indicated by
their dependence on the concentration profile~$\phi$.

The structure of the paper is the following: In the rest of the current section
we give a rough and qualitative description of the physics at work, exemplified
for a typical pattern of dense phases such as the one in Fig.~\ref{fig:device}b.
This specific example will be taken up again in the last section of the paper.
Section~\ref{sec:derivation} is dedicated to the detailed derivation of the
model proposed in this paper. It addresses in two distinct subsections the
modeling of the mixture behavior in the microfluidic channel and the
pervaporation through the membrane. All steps and assumptions made during the
derivation of the final equations are explained in the subsections. After the
derivation of the general equations we continue to explore several limit cases.
The section ends with a summary of the model equations and the underlying
assumptions. In Sec.~\ref{sec:proto_single} we focus on situations without phase
change, where we will study the smooth concentration profile of solute in the
dilute and the dense limits. In the following Sec.~\ref{sec:proto_several} we
then treat interfaces between different phases occurring in the mixture. The
motion of a single moving interface is described first, with the aim to extract
its dependence on the properties of the mixture. Then, we return to the
introductory example and analyze the thicknesses of several phase slabs. This
last analysis is done in a quasi\hyp{}stationary approximation. Both
sections~\ref{sec:proto_single} and \ref{sec:proto_several} employ numerically
obtained solutions of the model equations and their analytical approximations.
We finish with a summary and an outlook in Sec.~\ref{sec:summary}.
\subsection{The experimental parameters}

The experimental technique and the production of the microevaporator device are
detailed in Refs.~\cite{LenLonTabJoaAjd06,LenJoaAjd07,SalLen08}. In the
theoretical treatment in the present paper, we refer only to the most essential
properties of the device. They are displayed in Fig.~\ref{fig:device}. We use
three geometrical parameters, the length~$L$ of the PDMS~membrane, its
thickness~$e$ and the height~$h$ of the channel containing the solution. The
mixture enters the channel from a reservoir with a given solute
concentration~$\phi_L$. This concentration is fixed from outside by the
experimenter.

The mixture is driven out of equilibrium by the application of an air flow above
the PDMS membrane (thick arrows in Fig.~\ref{fig:device}a). The air flow
guarantees that the pervaporated water is constantly removed from the membrane
and that the permeation continues until the solute ``dries'' out. The
out\hyp{}of\hyp{}equilibrium situation is thus limited by the slow pervaporation. The
experimenter's second control parameter is the humidity in the air flow. In the
case of nearly saturated wet air no pervaporation takes place, because the
chemical potentials of water on both sides of the PDMS~membrane will become
equal. In general, the two external ``knobs,'' namely the reservoir
concentration~$\phi_L$ and the air humidity need not be constant but can be
changed according to a temporal protocol.

Due to the slow pervaporation process, a typical experiment takes hours to
days. According to Ref.~\cite{LenLonTabJoaAjd06}, the intrinsic time scales of
the device are in the range of seconds (for the evaporation and the mechanical
adjustment of the membrane) up to fifteen minutes (for the diffusion). We may
therefore hope that the mixture is not too ``far from equilibrium,'' which is a
necessary requirement for the following description.

During the evaporation, several quantities can be measured, such as the total
amount of mixture which passes through the device during the experiment. It
gives some information on the amount of solute in the channel and on the rate of
water pervaporation. Within the channel, more information on the mixture can be
gathered: If there are several phases in the mixture and if they can be
differentiated e.g.\ by their texture, then the position of the interfaces can
be tracked as a function of time. This has been done for the case of the
surfactant~AOT (docusate sodium salt) in water in Ref.~\cite{LenJoaAjd07} and
for a salt solution in Ref.~\cite{LenLonTabJoaAjd06}. More information is
obtained if the full density profile can be measured as a function of
time~\cite{LenLonTabJoaAjd06}. These types of observables are in the focus of
the present theoretical description.
\subsection{An intuitive example}
\label{sec:intuitive}

The central interest of the present theoretical description is the precise
functional forms of two coefficients, one of which is the inter\hyp{}diffusion
coefficient~$D(\phi)$ of solute and solvent, the other one $q(\phi)$ quantifies
the pervaporation of solvent through the PDMS~membrane. Both coefficients do
depend on the local concentration~$\phi(x,t)$ of solute. All thermodynamic and
kinetic properties of the mixture are implicitly contained in these two
coefficients. In particular, they can be expressed in terms of the chemical
potentials and of the kinetic properties of the mixture, both of which depend
on~$\phi$.

Suppose the situation in Fig.~\ref{fig:device}b, which has been explored
experimentally using the surfactant~AOT (docusate sodium salt) in
water~\cite{LenJoaAjd07}: The channel is filled by four different phases of the
mixture, which are ordered by their solute concentration. Three of them are
dense, indicated by different patterns in Fig.~\ref{fig:device}b, the fourth
phase is dilute. The pervaporation of water, quantified by introducing the
coefficient~$q(\phi)$ gives rise to a flow from the reservoir (long curved
arrows) which compensates the water loss at the PDMS~membrane. On its way
through the phases the water drags the solute with it. It thereby induces the
concentration of solute, and, at higher and higher concentration, the appearance
of the three dense phases which would be absent without flow.

The final concentration profile is thus a dynamic equilibrium between convective
and diffusive processes: Within each phase slab the concentration gradient gives
rise to diffusion which in turn tries to shrink the phases. While the diffusion
emerges together with the concentration gradients, the convective transport, or
``drag'' is reduced by the solute becoming dense. The solvent has to permeate
through the dense phases of solute before it reaches the membrane and
pervaporates. The denser the solute, the less solvent may pass. Furthermore, the
more the water flow tries to compress, or to ``squeeze'' the dense phases, the
harder it becomes to squeeze them even further. One or the other of the
mentioned mechanisms may remind the reader of terms such as ``osmotic
compressibility,'' ``diffusion,'' ``mobility,'' ``permeability,'' depending on
the liking and the experience of the reader. In fact, all these effects are
linked to one another, and we will provide the connection between some of these
terms in Sec.~\ref{sec:conn}. We thereby also show the equivalence of these
terms, in so far as they can expressed one by the other. In our model we abandon
the use of terms such as mobility, osmotic compressibility and permeation in
favor of the diffusion coefficient~$D(\phi)$ and the coefficient of
pervaporation~$q(\phi)$, which are expressed in terms of the chemical potential
of water and of the phenomenological coefficient for inter\hyp{}diffusion.

An important observation in the situation of Fig.~\ref{fig:device}b is the
following: As more and more solute is transported from the reservoir, all three
dense phases grow. It has been observed experimentally that the two interior
``sandwiched'' phases apparently grow at constant thickness. They loose as much
solute as they gain, while the last, the most dense phase continues to grow.
Keeping in mind the dynamic equilibrium between convective and diffusive
processes within each phase we therefore find in the thicknesses of these middle
phases a fingerprint of the diffusion coefficient and of the pervaporation
coefficient.

One may now ask whether a large diffusion coefficient will finally lead to a
thinner or to a thicker slab, as compared to one with a smaller diffusion
coefficient? On one hand, the diffusion coefficient scales the gradients of the
concentration profile which try to dissolve the phases. It should thus lead to
small thicknesses of the quenched dense phases. On the other hand, for fixed
values of the concentration at the phase interfaces, a large diffusion
coefficient will render the profiles flat and therefore lead to thick phases.
This may appear paradoxical on a first sight, but it only reflects the same
feedback mechanism as between flow of solvent and permeability. The resolution
is that we really need the full concentration profile and the velocity field in
order to understand the phase thicknesses.


\section{Detailed derivation of the model}
\label{sec:derivation}
In this section we derive a set of equations which model the evolution of a
binary system, reduced to one spatial dimension~$x$ along the evaporation
channel. As variables of state we use a concentration~$\phi(x,t)$ of solute and
a mixture velocity~$v^\vavg(x,t)$. Special focus will be put on their precise
meaning in the evolution equation that will be derived in the following
sections,
\begin{gather}
  \label{dxv_int}
  \partial_x v^\vavg = -\vol_w q(\phi), \\
  \label{dtphi_int}
  \partial_t\phi = -\partial_x\bigl[\phi v^\vavg - D(\phi) \partial_x\phi\bigr].
\end{gather}
Here, $\vol_w$ denotes the specific mass of water; $\partial_t$ and $\partial_x$
are partial derivatives with respect to time and space. The physical content of
the equations is the independent balance of solvent and solute mass. The variables
are chosen such that $q(\phi)$~reflects the loss of solvent (water) via
pervaporation, and $D(\phi)$~is a state\hyp{}dependent inter\hyp{}diffusion coefficient,
which contains information on the thermodynamic and dynamic properties of the
mixture. The expressions for $D(\phi)$ and $q(\phi)$ will be provided by
equations \eqref{diffcoeff} and \eqref{qmu} below.

\subsection{Diffusion model for the binary solution}
\label{sec:mixture}
The aim here is to state clearly all underlying assumptions which lead to the
model equations \eqref{dxv_int}~and \eqref{dtphi_int}, starting from mass
conservation. This procedure will allow to point out the restrictions of the
model. We employ the theory of \emph{linear out\hyp{}of\hyp{}equilibrium thermodynamics}
as summarized in the book by de~Groot and Mazur~\cite{GroMaz84}. We try to adopt
the notation used there, also for an easier comparison with the results by
Peppin~\emph{et\,al.}~\cite{PepEllWor05} for the link to sedimentation,
ultra\hyp{}filtration, and reverse osmosis.
\subsubsection{Balance equations for binary mixtures}

The balance equations for the mass densities $\rho_s(\bfr, t)$ and $\rho_w(\bfr,
t)$ of solute (index~$s$) and solvent (index~$w$) read
\begin{equation}
  \label{3D:evol}
  \partial_t \rho_k = -\div(\bfj_k), \quad\text{with $k \in \{s, w\}$.}
\end{equation}
The mass current densities~$\bfj_k$ are split into an advected and a diffusive
part,
\begin{gather}
  \label{3D:js}
  \bfj_s = \rho_s\bfv_s = \rho_s \bfv + \bfJ_s, \\
  \label{3D:jw}
  \bfj_w = \rho_w\bfv_w = \rho_w \bfv + \bfJ_w.
\end{gather}
The $\bfJ_i$ denote the ``diffusive currents'', simply defined by the difference
$\rho_i (\bfv_i - \bfv)$. A common choice for the mixture velocity is the
mass\hyp{}average of the individual constituent velocities~$\bfv_k$, weighted by the
mass fractions~$c_k$,
\begin{gather}
  \label{v}
  \bfv := \sum_i c_i \bfv_i\\
  \text{with}\quad
  c_i := \frac{\rho_i}{\rho} \quad\text{and}\quad \rho:= \sum_k \rho_k.
\end{gather}
This choice makes the two diffusive currents opposite to each other,
\begin{equation}
  \label{JwJs}
  \bfJ_w = -\bfJ_s.
\end{equation}

Formally, the mass densities are taken to be functions of the local
thermodynamic variables of state as~$\rho_i(T, P, c_s)$. The use of local
temperature and pressure fields shows that here and in the following we will
\emph{assume local thermodynamic equilibrium}. The evolution equations
\eqref{3D:evol} are accompanied by the following constraint, which is an
immediate consequence of the chemical potentials being intensive quantities (or
of the Euler relation),
\begin{equation}
  \label{volfrac}
  1 = \vol_s\rho_s + \vol_w\rho_w.
\end{equation}
The specific volumes per mass, $\vol_i$, are defined as the derivatives of the
chemical potentials per mass, $\chpot_i(T, P, c_s)$, with respect to the pressure,
\begin{gather}
  \label{voldef}
  \vol_i(T, P, c_s) := \left.\frac{\partial\chpot_i}{\partial P} \right|_{T,c_s}. \\
 \chpot_i(T, P, c_s) := \frac{\mu_i(T, P, c_s)}{m_i}.
\end{gather}
Here, $\mu_i$ is the chemical potential per molecule, and $m_i$ the mass of such
a molecule. The derivative in Eq.~\eqref{voldef} equals the change of volume
induced by adding an infinitesimal amount of mass of type~$i$ upon keeping
temperature, pressure and all remaining masses constant.
Equation~\eqref{volfrac} can be used to introduce the \emph{volume fractions}
$\phi_i(T,P,c_s) := \vol_i\rho_i$. Below, equation~\eqref{volfrac} will also be
the point of entrance for the notion of \emph{simple mixtures} and of
\emph{incompressibility} of the constituents. Apart from Eq.~\eqref{volfrac},
another consequence of the intensivity of chemical potentials is the following
relation between the derivatives of the two chemical potentials with respect to
the mass fraction of solvent, $c_s$,
\begin{equation}
  \label{chempoteq}
  c_s \frac{\partial\chpot_s}{\partial c_s}\biggr|_{T,P}
  = - (1-c_s) \frac{\partial\chpot_w}{\partial c_s}\biggr|_{T,P}.
\end{equation}
This equation establishes the standard logarithmic behavior of the chemical
potentials in the limits $c_s\to 0$ and $c_s\to 1$.

Together, equations~\eqref{volfrac} and~\eqref{chempoteq} reduce the number of
relevant chemical potentials to one, as one can always be expressed by the
other. This fact is important for the number of kinetic coefficients within the
theory of linear non\hyp{}equilibrium thermodynamics. Applying this framework in the
present description, the diffusive current~$\bfJ_s$ is taken to be proportional
to the gradients of chemical potentials. The proportionality factor is called
the \emph{phenomenological coefficient}~$\PhCoeff$. Using the relation of
Gibbs-Duhem, only the difference of chemical potentials can occur, and we
receive for the diffusive current
\begin{equation}
  \label{Js}
  \bfJ_s = - \PhCoeff\:\grad(\chpot_s - \chpot_w).
\end{equation}
We do not present the details of the derivation of this equation here, as it
implies a rather tedious consideration of all possible symmetries in the entropy
production. Instead, we refer to Eq.~(IV.15) in the book by de~Groot and
Mazur~\cite{GroMaz84} or to the summary in the paper by
Peppin~\emph{et\,al.}~\cite{PepEllWor05}. Note that external forces such as
gravity cancel out and do not appear in equation~\eqref{Js}. This fact is due to
the chemical potentials being defined as energy per mass rather than per number
of particles. Furthermore, we have omitted a term proportional to the
temperature gradient, as from now on we take the temperature as uniform and
constant. We will omit it also from the argument lists of thermodynamic
functions. The \emph{phenomenological coefficient}~$\PhCoeff$ may in principle
depend on all thermodynamic variables of state as well as on their derivatives,
and even on the mixture velocity~$\bfv$ and its derivatives. However, if the
assumption of local thermodynamic equilibrium is well satisfied, the dependence
on the mixture velocity field should be negligible, such that we use a function
$\PhCoeff(P, c_s)$ for the moment. The chemical potentials are also functions of
the two thermodynamic parameters: $\chpot_s(P, c_s)$ and $\chpot_w(P, c_s)$.
This functional dependence allows to separate the influence of pressure
gradients from concentration gradients in Eq.~\eqref{Js},
\begin{equation}
  \label{Js_grads}
  \bfJ_s = - \PhCoeff\:\Bigl[(\vol_s - \vol_w)\grad P +
  \frac{\partial(\chpot_s-\chpot_w)}{\partial c_s}\Bigr|_{T,P}\grad c_s
  \Bigr].
\end{equation}
With the aid of relation~\eqref{chempoteq} the diffusive second term can be
rephrased using either of the chemical potentials,
\begin{equation}
  \label{diff_equiv}
  \frac{\partial(\chpot_s-\chpot_w)}{\partial c_s}\Bigr|_{T,P}
  = \frac{1}{1-c_s} \frac{\partial\chpot_s}{\partial c_s}\Bigr|_{T,P}
  = - \frac{1}{c_s} \frac{\partial\chpot_w}{\partial c_s}\Bigr|_{T,P}.
\end{equation}
Equation~\eqref{Js_grads} has been used by
Peppin~\emph{et\,al.}~\cite[Eq.~(28)]{PepEllWor05} to demonstrate that Fick's
and Darcy's laws are limiting cases of the same equation. Below, we will use it
to define the diffusion coefficient and to make the connection between the
different methods micro\hyp{}pervaporation, sedimentation, and ultra\hyp{}filtration.
\subsubsection{Reduction to 1D}

\begin{figure}[hbt]
  \centering
  \includegraphics{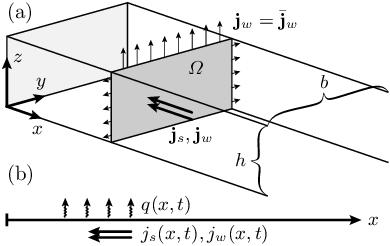}%
  \caption{Visualization of the averaging used for reducing the balance
  equations to one dimension. (a)~The area of integration~$\Omega$ in three
  dimensions. (b)~The one\hyp{}dimensional representation.}%
  \label{fig:slice}%
\end{figure}

We shall now disregard the details of the flow pattern in the channel and reduce
the three\hyp{}dimensional evolution equations~\eqref{3D:evol} to one dimension. We
do this by averaging over the two spatial dimensions perpendicular to the
channel axis, as depicted in Fig.~\ref{fig:slice}. For a given uniform
height~$h$ and width~$b$ of the channel we obtain from integration over a slice
$\Omega$ with area $|\Omega| = bh$ the one\hyp{}dimensional counterparts of
\eqref{3D:evol},
\begin{align}
  \label{dtrho_s_1d}
  \partial_t \rho_s &= -\partial_x j_s \\
  \label{dtrho_w_1d}
  \partial_t \rho_w &= -\partial_x j_w - q.
\end{align}
The new densities and currents turn out to be
\begin{gather}
  \rho_k(x,t) := \frac{1}{|\Omega|} \int_\Omega dy\,dz\: \rho_k(x,y,z,t), \\
  j_k(x,t) := \frac{1}{|\Omega|} \int_\Omega dy\,dz\: \bfe_x\dotp\bfj_k(x,y,z,t).
\end{gather}
The water loss due to pervaporation through the PDMS membrane is all included
in~$q(x,t)$. It represents the integral of water current in normal direction
over the boundary of the channel,
\begin{align}
  \label{defq}
  q(x,t) &:= \frac{1}{|\Omega|} \int_{\partial\Omega}
  \bfN\dotp\bfj_w(x,y,z,t) \\
  \label{defq_approx}
         &\approx \frac{1}{b\,h} \int_0^b \bfN\dotp\bfj_w(x, y, h, t) \,dy.
\end{align}
The pervaporation has a dominant contribution from the upper boundary where the
thin layer of PDMS allows a steady water pervaporation. In addition to
equations~\eqref{dtrho_s_1d} and \eqref{dtrho_w_1d} we expect an averaged
version of constraint~\eqref{volfrac} to hold. In the same manner as above
in Eqs.~\eqref{3D:js} and \eqref{3D:jw} we may subdivide the total currents into
advective and diffusive parts.
%

\subsubsection{Incompressibility and simplicity of mixtures}

We now assume both constituents of the mixture to be individually
incompressible, having constant specific volumes~$\vol_i$. This assumption will
allow to reduce the set of three equations~\eqref{dtrho_s_1d},
\eqref{dtrho_w_1d}, and \eqref{volfrac} to only two equations. The assumption,
however, is quite strong as it implies the mixture to be both incompressible and
simple,
\begin{equation}
  \frac{\partial 1/\rho}{\partial P}\biggr|_{T,c_s} = 0, \qquad
  \frac{\partial 1/\rho}{\partial c_s}\biggr|_{T,P} = \vol_s - \vol_w =
  \textit{const.}
\end{equation}
The term \emph{simple mixture} here means that the total mass density is a
linear function of the mass fraction. It implies linear relations between mass
density, mass fraction and volume fraction. We will therefore use the terms
``density'', ``fraction'' and ``concentration'' interchangeably. Simple mixtures
are for example glycerol in water~\cite{JosHuaHu96} or rigid spheres in an
incompressible solvent of much smaller particles. The constraint~\eqref{volfrac}
allows to write one density as a function of the other, $\rho_w =
\rho_w(\rho_s)$. The evolution equation~\eqref{dtrho_w_1d} for $\rho_w(\rho_s)$
then becomes
\begin{equation}
  \label{eq:full_dxv}
  \Bigl(\rho_w(\rho_s) - \frac{\partial\rho_w}{\partial\rho_s}\rho_s\Bigr) \partial_x v
    + \partial_x J_w - \frac{\partial\rho_w}{\partial\rho_s} \partial_x J_s
    = -q.
\end{equation}
This equation can be understood as a differential equation for the velocity
field, similar to the one we are looking for~\eqref{dxv_int}. However, the
occurrence of the diffusive currents~$J_s$ and $J_w$ makes it unfavorably
dependent on the concentration variable, see equations~\eqref{Js_grads} and
\eqref{JwJs}. A~more convenient equation is obtained by using from the beginning
the volume\hyp{}averaged mixture velocity
\begin{equation}
  v^\vavg := \sum_i \phi_i v_i
\end{equation}
instead of the mass\hyp{}averaged one. The total currents are then written with
modified diffusive currents,
\begin{equation}
  j_k = \rho_k v^\vavg + J^\vavg_k,\quad k\in\{s,w\}.
\end{equation}
With the volume\hyp{}averaged mixture velocity, and by setting the specific volumes
constant, one obtains as relation between the new diffusive currents,
\begin{equation}
  J^\vavg_w = -\frac{\vol_s}{\vol_w} J^\vavg_s
\end{equation}
instead of equation~\eqref{JwJs}. This property makes the diffusive currents
vanish from the differential equation for~$v^\vavg(x,t)$, which now comes close
to the form we seek~\eqref{dxv_int},
\begin{equation}
  \label{dxv}
  \partial_x v^\vavg = -\vol_w q(P,\phi).
\end{equation}

In terms of the volume\hyp{}averaged mixture velocity, and with constant~$\vol_s$,
the volume fraction of solute evolves as
\begin{equation}
  \partial_t \phi_s = - \partial_x\Bigl[\phi_s v^0 + \bigl(\vol_w\phi_s +
  \vol_s(1{-}\phi_s)\bigr)\,J_s\Bigr].
\end{equation}
Or, omitting the index $s$ and collecting also equations~\eqref{Js_grads} and
\eqref{diff_equiv}, we finally arrive at
\begin{equation}
  \label{dtphi}
  \partial_t \phi = -\partial_x\Bigl[\phi v^0
   - K(P, \phi) \partial_x P
   - D(P, \phi) \partial_x\phi
   \Bigr],
\end{equation}
with the shortcuts
\begin{gather}
  \label{presscoeff}
  K(P, \phi) = [\vol_s + \phi(\vol_w-\vol_s)] \: \PhCoeff(P,\phi) \: (\vol_s-\vol_w), \\
  \label{diffcoeff}
  D(P, \phi) = -\frac{[\vol_s + \phi(\vol_w-\vol_s)]^2}{\vol_w\phi} \: \PhCoeff(P,\phi) \:
   \frac{\partial\chpot_w}{\partial\phi}\Bigr|_{T,P}.
\end{gather}
Equations \eqref{dxv}~and \eqref{dtphi} with the coefficients $q$~and $D$
from~\eqref{defq} and \eqref{diffcoeff} come close to the model we announced in
the introduction. They nicely separate the effect of pervaporation, which is all
contained in the coefficient~$q$, from the bulk thermodynamic and kinetic
properties of the mixture. These properties are implicitly contained in the
functional dependence of $\PhCoeff$ and~$\chpot_w$ on $P$ and on~$\phi$.


\subsubsection{The role of the pressure gradient}
\label{sec:pressgrad}

With equations \eqref{dxv}~and \eqref{dtphi} we have found two equations for yet
three variables $v^\vavg$, $\phi$, and $P$. We thus have to seek for a closure
of the above set of equations. The contribution by the pressure requires some
further attention. The pressure occurs as an argument in several functions, but
also its gradient adds a contribution to the diffusive current in
Eq.~\eqref{dtphi}. This gradient appears also in the (compressible) Stokes
equation. In principle, it would have been possible to assume \emph{local
mechanical equilibrium} already in the 3D~theory, such as the $x$-component of
the compressible Stokes equation,
\begin{equation}
  \label{Stokes}
  0 = -\partial_x P + \eta \Delta v_x + (\lambda + \eta/3) \partial_x \div\bfv
\end{equation}
and then reduce it to one dimension together with fitting boundary conditions.
The shear and bulk viscosities are denoted by $\eta$~and $\lambda$,
respectively. However, this procedure raises fundamental difficulties: We showed
above that the natural description of pervaporation of incompressible mixtures
involves the volume\hyp{}averaged velocity~$v^\vavg$. The natural formulation of the
Stokes equation uses the mass\hyp{}averaged velocity. Only for the special case
$\vol_s = \vol_w$ both velocities do coincide---but in this very case the term
with $\partial_x P$ vanishes due to $K=0$. When expressing $\bfv$ by
$\bfv^\vavg$, the term with $\partial_x P$ in Eq.~\eqref{dtphi} will contribute
to both other terms in the expression for the diffusive current
in~\eqref{dtphi}. We do even expect a coupling of velocity and density
variables, as well as a second derivative $\partial_x^2 \phi$ due to the second
derivatives in the Stokes equation~\eqref{Stokes}.

The precise analysis of this splitting is beyond the scope of the present paper.
Here, we have to restrict the validity of the model to cases where the
\emph{influence of the pressure gradient is negligible}. An
estimation~\cite{footnote3} indicates that this restriction is indeed weak: In
the dilute limit, using values from the experiment~\cite{LenLonTabJoaAjd06}, the
term with the pressure gradient in Eq.~\eqref{dtphi} is several magnitudes
smaller than the diffusion term with the concentration gradient. In what follows
we therefore use the simplistic model with a homogeneous pressure variable
(which we omit in the notation from now on):
\begin{gather}
  \label{dxv_final}
  \partial_x v^\vavg = -\vol_w q(\phi), \\
  \label{dtphi_final}
  \partial_t \phi = -\partial_x\Bigl[\phi v^0
   - D(\phi) \partial_x\phi
   \Bigr].
\end{gather}
with the \emph{evaporation coefficient}~$q(\phi)$ and the \emph{diffusion
coefficient}~$D(\phi)$, both depending only on the volume fraction~$\phi$.

As we take the microevaporation channels to be horizontal, our assumption of
setting the pressure gradient to zero coincides with the assumptions employed by
Peppin~\emph{et\,al.}~\cite{PepEllWor05} and also in the context of
sedimentation~\cite{RusSavSch89} where the pressure gradient balances
gravitational forces. It therefore opens the possibility to consistently compare
with their model equations, see Sec.~\ref{sec:conn}.

\subsubsection{Boundary conditions}

Equations \eqref{dxv_final} and \eqref{dtphi_final} are to be accompanied by an
appropriate number of boundary conditions. For the hyperbolic differential
equation~\eqref{dxv_final} we require one, for the parabolic
equation~\eqref{dtphi_final} we need two boundary conditions. At the end of the
channel, both the mixture velocity and the total solute current vanish,
\begin{gather}
  \label{bc:v}
  v^\vavg(0, t) = 0,\\
  \label{bc:dxphi}
  \partial_x \phi(0, t) = 0.
\end{gather}
As a third boundary condition we use a fixed concentration at the inlet of the
channel, reflecting the concentration in the reservoir,
\begin{equation}
  \label{bc:phi}
  \phi(L, t) = \phi_L.
\end{equation}
The model equations could equally handle a time\hyp{}dependent reservoir
concentration~$\phi_L(t)$. In a setup with a fixed total amount of solute in the
channel, it is mathematically also possible to replace the latter boundary
condition by an integral condition setting $\int_0^L\phi(x,t)\,dx$ to a given
value.
\subsubsection{Moving interfaces between phases}
\begin{figure}[bt]
  \centering
  \includegraphics{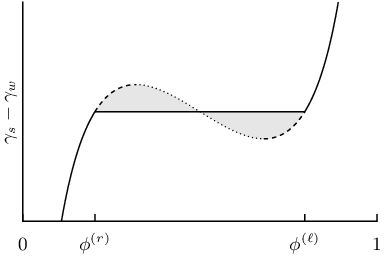}\\
  \includegraphics{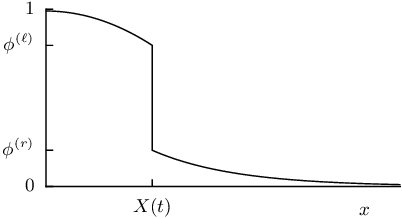}%
  \caption{Sketch of the modeling of sharp interfaces between different phases.
  The concentration values left and right of the interface are fixed by a
  Maxwell construction on the chemical potential (upper panel), the two regions
  shaded in gray have equal area.}%
  \label{fig:interface}%
\end{figure}%

In the introductory example we mentioned that the positions of phase interfaces
and their velocities are convenient observables in the pervaporation
experiments. Such experimental observations have been reported in
Ref.~\cite{LenLonTabJoaAjd06} and in more detail in Ref.~\cite{LenJoaAjd07} for
a solution of the surfactant~AOT (docusate sodium salt) in water. The essential
observation in the latter reference is the \emph{decrease} of the interface
velocity with time, which will be investigated in more detail below in
section~\ref{num:moving}. A quantitative analysis of this decrease can give
insight into the mechanisms at work, especially on the relative importance of
diffusion and advection caused by pervaporation.

Throughout this paper, we model the interfaces within the framework of local
equilibrium thermodynamics, which implies that the interfaces are infinitely
sharp and that for each such interface there are two unique concentration values
at its left\hyp{}hand and its right\hyp{}hand sides. These values are determined by a
Maxwell construction on the local chemical potential function.
Figure~\ref{fig:interface} sketches this construction and the resulting
concentration profile around the interface. The fixed concentration variables
serve as two Dirichlet boundary conditions for the evolution
equation~\eqref{dtphi_final},
\begin{equation}
  \label{int:phifixed}
  \phi\bigl(X^\lef, t\bigr) = \phi^\lef, \quad\phi\bigl(X^\rig, t\bigr) =
  \phi^\rig.
\end{equation}
The superscripts $\lef$ and $\rig$ indicate the left and right limits towards
the interface position~$X(t)$.

The conservation of solute mass, which is valid also across the interface,
provides us with a continuity condition at the interface. If the boundary does
not move, the solute currents at the left\hyp{}hand and at the right\hyp{}hand sides of
the interface must coincide. Any possible difference between these incoming and
outgoing currents must lead to a growth of one phase and thus to the movement of
the interface position~$X(t)$. Its velocity times the concentration jump at the
surface must equal the difference between the currents,
\begin{gather}
  \label{int:dotX}
  \dot X(t) = \frac{j_s\bigl(X^\lef, t\bigr) - j_s\bigl(X^\rig, t\bigr)}{\phi^\lef - \phi^\rig}.
\end{gather}

We further require continuity conditions for the velocity field. As we did in
the derivation of the bulk equations, we also here want the mixture velocity to
reflect only the influence of the pervaporation. As we do not expect the
pervaporation to diverge near phase interfaces, the velocity field should be
continuous across such an interface,
\begin{equation}
  \label{int:continv}
  v^\vavg(X^\lef, t) = v^\vavg(X^\rig, t).
\end{equation}
This continuity equation is consistent with the conservation of solvent mass
across the interface.

\subsection{Pervaporation through the PDMS membrane}
\label{sec:pervaporation}
The pervaporation of water through the PDMS layer can be described on the same
level as the advective and diffusive currents in the evaporation channel. We now
focus on the binary mixture of water and PDMS, as solutes cannot enter the PDMS
domain. The notation is such that the properties inside the PDMS membrane will
be denoted by an overbar. As we did above, we also here split the total current
density~$\barbfj_w$ of water into convective and diffusive terms. This time it
is more convenient to take the velocity of PDMS, which is zero, as the reference
velocity of the mixture. The mass current density of water in the layer then
becomes
\begin{equation}
  \barbfj_w = \frac{1}{1-\barc_w} \barbfJ_w,
\end{equation}
where $\barc_w$ is the mass fraction of water \emph{inside the layer}, and
$\barbfJ_w$ is the corresponding diffusive current. In complete analogy to
equation~\eqref{Js}, the diffusive current can be written in terms of a
phenomenological coefficient~$\barL$,
\begin{equation}
  \label{evap_one}
  \barbfj_w = -\frac{\barL(P, \barc_w)}{1 - \barc_w} \grad (\barchpot_w - \barchpot_\PDMS).
\end{equation}
with the chemical potential of PDMS denoted by~$\barchpot_\PDMS$. We understand
the pervaporation as a diffusive process and again omit the pressure\hyp{}dependence
of the chemical potentials. Then, the water current is proportional to the
gradient of water concentration. The diffusion process is further assumed to be
sufficiently established that the water current does not vary within the
membrane. In the dilute limit, where the chemical potential exhibits a
logarithmic behavior,
\begin{equation}
  \label{barchpotw_dilute}
  \barchpot_w(T, P, \barc_w) = \barchpot_w^\ast(T, P) + \frac{kT}{m_w}\,\ln
  \barc_w,
\end{equation}
the uniform water current then corresponds to a constant gradient of water
concentration inside the membrane. The water current through the membrane is
then proportional to the concentration difference across the membrane of
thickness~$e$,
\begin{equation}
  \label{evap_two}
  \bfN\dotp\barbfj_w =
  \frac{\barc_w(z{=}h{+}e) - \barc_w(z{=}h)}{e}
  \lim_{\barc_w\to0}\frac{\barL(P, \barc_w)}{(1 - \barc_w)^2}
  \frac{\partial\barchpot_w}{\partial \barc_w}.
\end{equation}

To make the connection with the conditions in the adjacent channels of the
mixture and of the air flow, we assume that the chemical potentials at both
sides of both boundaries of the PDMS~membrane have matching values,
\begin{align}
  \barchpot_w(z{=}h) = \chpot_w \quad\text{and}\quad
  \barchpot_w(z{=}h{+}e) = \chpot_w^\air.
\end{align}
The evaporation coefficient $q$~can now be expressed as a function of the
chemical potential $\chpot_w(\phi)$ in the mixture channel. From equations
\eqref{evap_two}, \eqref{barchpotw_dilute}, and its
definition~\eqref{defq_approx} we finally obtain the following dependence of the
pervaporation coefficient on the volume fraction,
\begin{equation}
  \label{qmu}
  q(\phi) = \frac{1}{\vol_w} \frac{\barD}{eh} \biggl[
     \exp\Bigl\{\frac{m_w}{kT}\chpot_w(\phi)\Bigr\} -
     \exp\Bigl\{\frac{m_w}{kT}\chpot_w^\air\Bigr\}
  \biggr].
\end{equation}
We have collected several of the above constants in the shortcut~$\barD$,
\begin{equation}
  \barD = \vol_w \exp\Bigl\{-\frac{m_w}{kT}\barchpot_w^\ast\Bigr\}
  \lim_{\barc_w\to0} \barL\, \frac{\partial\barchpot_w}{\partial\barc_w}.
\end{equation}
Expression ~\eqref{qmu} contains the further assumption that the chemical
potential directly at the interface between mixture channel and PDMS layer is
the same as the average chemical potential in the mixture which we obtained from
the reduction to one spatial dimension. In the present framework of a
one\hyp{}dimensional description, there is no alternative to this approximation.

At the present point we have completed the derivation of the model, consisting
of the two differential equations~\eqref{dxv_final} and \eqref{dtphi_final} for
the volume fraction~$\phi$ and the volume\hyp{}averaged velocity
field~$v^\vavg$, together with the concentration\hyp{}dependent diffusion
coefficient~\eqref{diffcoeff} and the evaporation coefficient~\eqref{qmu}.

\subsection{Connection with sedimentation and ultra-filtration}
\label{sec:conn}
We now make the connection between the two coefficients $D(\phi)$ and $q(\phi)$
and other pairs of coefficients. This will allow to compare different techniques
of measuring chemical potentials and phenomenological coefficients and to verify
the consistency of the outcomes. We will not make use of these connections in
the present paper. The equations in this section can rather be seen as a
convenient reference for the reader working on different techniques. However,
the comparisons give some more intuition on the variables and coefficients which
are used. Above, in equations~\eqref{diffcoeff} and \eqref{qmu} we have already
found the link to the pair $\chpot(\phi)$ and $\PhCoeff(\phi)$. We continue by
expressing also the coefficients of sedimentation and of permeability in terms
of $\chpot(\phi)$ and $\PhCoeff(\phi)$. Each of these two coefficients forms
together with the diffusion coefficient a pair which allows to determine
$\chpot(\phi)$ and $\PhCoeff(\phi)$ and thus to make the link to our original
coefficients $D(\phi)$ and $q(\phi)$.
\subsubsection{Sedimentation factor and osmotic compressibility factor}

In the context of sedimentation of colloids, Russel, Saville, and
Schowalter~\cite{RusSavSch89} express the thermodynamic and dynamic properties
in terms of the \emph{sedimentation factor} $K^\RSS(\phi)$ and the \emph{osmotic
compressibility factor}~$Z^\RSS(\phi)$. Both can be introduced by rewriting
Eq.~\eqref{dtphi} as follows~\cite[compare Eq.~12.5.6]{RusSavSch89},
\begin{equation}
  \label{RSS}
  \partial_t\phi = -\div\Bigl(\phi\bfv^\vavg + \bfU_0 \phi K^\RSS(\phi) -
  D(\phi)\bfnabla\phi\Bigr)
\end{equation}
with the diffusion coefficient
\begin{equation}
  D(\phi) = D_0 K^\RSS(\phi) \frac{d}{d\phi}\bigl[\phi Z^\RSS(\phi)\bigr]
\end{equation}
and with the shortcuts
\begin{equation}
  D_0 := \frac{kT}{6\pi\eta\,a},\quad
  \bfU_0 := -\frac{2 a^2}{9\eta}\Bigl(\frac{1}{\vol_w}-\frac{1}{\vol_s}\Bigr)\bfg
\end{equation}
with $\eta$~denoting the viscosity of pure solvent, $a$~the radius of a
colloidal sphere, and $\bfg$~the acceleration by gravity. The volume\hyp{}averaged
velocity in Eq.~\eqref{RSS} vanishes by construction of the sedimentation
experiment in a closed container. In the comparison with Eq.~\eqref{dtphi} the
remaining two terms thus lead us to the sedimentation coefficient~$K^\RSS$ given
as
\begin{equation}
  \label{KRSS}
  K^\RSS(\phi) = \frac{|\bfnabla P|}{\rho g} \frac{9\eta}{2a^2} \vol_s^2
    \Bigl(1 + \phi\frac{\vol_w{-}\vol_s}{\vol_s}\Bigr)^2
    \frac{\PhCoeff(\phi)}{\phi}.
\end{equation}
The osmotic compressibility factor, originally defined by the osmotic pressure
of solute~$\Pi(\phi)$ can equivalently be expressed in terms of the chemical
potential of water,
\begin{equation}
  Z^\RSS(\phi) = \frac{3}{4\pi\,a^3} \frac{\Pi(\phi)}{\phi kT}
  = \frac{3}{4\pi\,a^3} \frac{\chpot_w(0) - \chpot_w(\phi)}{\vol_w \phi kT}.
\end{equation}

Peppin, Elliott, and Worster~\cite{PepEllWor05} define the \emph{sedimentation
coefficient}~$S^\PEW$ in a marginally different way. Their definition leads to
the following variant of the transport equation~\eqref{RSS} (see Eq.~(32) of
Ref.~\cite{PepEllWor05}),
\begin{equation}
  \label{PEW}
  \partial_t\phi = -\div\Bigl(\phi\bfv^\vavg + \bfg\phi K^\PEW(\phi) - D(\phi)\bfnabla\phi\Bigr)
\end{equation}
In their treatment, the pressure gradient is assumed to balance the force
density by gravitation, $\bfnabla P = \rho \bfg$. In our notation of
Eq.~\eqref{presscoeff}, their sedimentation coefficient reads
\begin{equation}
  S^\PEW(\phi) = \frac{K(\phi)}{\phi\vol_w} \Bigl(1 +
  \phi\frac{\vol_w-\vol_s}{\vol_s}\Bigr).
\end{equation}
%

\subsubsection{Permeability}
\label{sec:permeab}

For the coefficient of permeability~$k^\PEW$ Peppin~\emph{et\,al.} give the
result (Eq.~(37) of Ref.~\cite{PepEllWor05})
\begin{equation}
  \label{kPEW}
  \frac{k^\PEW}{\eta} = \frac{\PhCoeff(\phi)\vol_s}{\phi^2 \vol_w^2} \Bigl(1 +
  \phi\frac{\vol_w-\vol_s}{\vol_s}\Bigr)^2.
\end{equation}
Except for the dilute limit, the phenomenological coefficient~$\PhCoeff(\phi)$
can be understood as a direct measure of the permeability of the solute with
respect to the solvent. This will be useful below when choosing a model function
for~$\PhCoeff(\phi)$. In particular, Eq.~\eqref{kPEW} helps choosing the value
of $\PhCoeff(1)$, which has influence on the qualitative behavior of the
solutions, see the discussion of Fig.~\ref{fig:L}. In the opposite limit,
Eq.~\eqref{KRSS} indicates that $\PhCoeff(\phi) \propto \phi$ for small~$\phi$.
This last requirement leads to a finite non\hyp{}zero diffusion coefficient in the
dilute limit, see below in Fig.~\ref{fig:sharp1}.

\subsection{Summary of the model equations, and their preparation for a numerical implementation}
We here summarize the model equations \eqref{dxv_int}, \eqref{dtphi_int},
\eqref{diffcoeff}, and \eqref{qmu} for their further use in the following
sections. We repeat their boundary conditions and the condition for the
interface movement. We also reformulate the model equations in such a way that
they may be discretized and implemented more easily. The dimensionless
equations of evolution \eqref{dxv_int} and \eqref{dtphi_int} read
\begin{gather}
  \label{dxv_num}
  \partial_{x'}v^{0\prime} = - q'(\phi), \\
  \label{dtphi_num}
  \partial_{t'}\phi
   = - \partial_{x'} j'
   = - \partial_{x'} \Bigl[
    v^{0\prime}\phi - \frac{D'(\phi)}{\Pe} \partial_{x'}\phi\Bigr].
\end{gather}
They describe the evolution of an incompressible binary mixture in one
effective spatial dimension under the further assumption of homogeneous
temperature and pressure. The variables carrying a prime are made dimensionless
by the following scales for length ($x$), time ($t$) and the mixture velocity
($v^\vavg$). The volume fraction $\phi$ needs no further rescaling. The scales
have been chosen such that all dimensionless variables are of the order unity
as far as possible,
\begin{gather}
  x = L x', \quad
  t = t' / \bigl(\vol_w q(0)\bigr), \quad
  v^\vavg = L\vol_wq(0)\, v^{0\prime}, \\
  q(\phi) = q(0)\, q'(\phi), \quad D(\phi) = D(0)\, D'(\phi), \\
  \Pe = \frac{L^2 \vol_w q(0)}{D(0)}.
\end{gather}
Here, the Peclet number~$\Pe$ is a global Peclet number of the system. The
scales for the diffusion coefficient and the coefficient of evaporation, which
are simply the coefficients evaluated at zero concentration density ($D(0)$ and
$q(0)$) are chosen such that $\Pe$~presents an upper bound for the true (local)
Peclet number~$\Peloc(x,t)$ which can only be determined a posteriori from the
solution. In numerical solutions we have found the local Peclet number to be
always maximal at the entrance of the channel and unity at the end. However, it
does not need to be monotonous.

The dimensionless version $q'(\phi)$ of the pervaporation coefficient is, according
to the expression we found in~\eqref{qmu},
\begin{equation}
  \label{qmu_num}
  q'(\phi) = \frac%
  {\exp\Bigl\{\frac{m_w}{kT}\bigl(\chpot_w(\phi)-\chpot_w^\air\bigr)\Bigr\} - 1}
  {\exp\Bigl\{\frac{m_w}{kT}\bigl(\chpot_w(0)   -\chpot_w^\air\bigr)\Bigr\} - 1}.
\end{equation}
It depends on the volume fraction~$\phi$ only via the chemical potential of
water~$\chpot_w(\phi)$, all other parameters are constants or external driving
parameters. The dimensionless diffusion coefficient reads, according to
equation~\eqref{diffcoeff},
\begin{equation}
  \label{diffcoeff_num}
  D'(\phi) = \frac{[1 + \phi(\vol_w/\vol_s -1)]^2}{\phi} \:
   \frac{\PhCoeff(\phi)\: \partial_\phi\chpot_w(\phi)}
   {\partial_\phi\chpot_w(0)\: \lim\limits_{\varphi\to0}\PhCoeff(\varphi)/\varphi}.
\end{equation}
It obtains it dependence on $\phi$ from~$\chpot_w(\phi)$ and from the
phenomenological coefficient $\PhCoeff(\phi)$ for inter\hyp{}diffusion.

The boundary conditions \eqref{bc:v}, \eqref{bc:dxphi}, \eqref{bc:phi} read in
their rescaled version,
\begin{align}
  v^{\vavg\prime}(x', t') = 0 &\quad\text{at $x'=0$},\\
  \partial_{x'}\phi(x', t') = 0 &\quad\text{at $x'=0$},\\
  \phi(x', t') = \phi_L &\quad\text{at $x'=1$},
\end{align}
as well as the boundary/continuity conditions at moving phase interfaces
\eqref{int:phifixed} and \eqref{int:continv}. The velocity \eqref{int:dotX} of
an interface becomes
\begin{multline}
  \label{dtX_num}
  \partial_{t'} X'(t) = v^{0\prime}(X',t) \\ +
  \frac{-D(\phi^\lef)\partial_{x'}\phi(X^{\prime\lef},t) +
         D(\phi^\rig)\partial_{x'}\phi(X^{\prime\rig},t)}{\Pe\:[\phi^\lef - \phi^\rig]}
\end{multline}

The following sections will require solutions for the model equations, partly
for arbitrary functions $q(\phi)$ and $D(\phi)$. Such solutions are available
only by numerical methods. For the numerical discretization we transform
equations \eqref{dxv_num} and \eqref{dtphi_num} into time\hyp{}dependent systems of
coordinates $\xi(x',t')$ which range from $0$~to $1$ in each phase. $x'(\xi,t')$
interpolates linearly between the (moving) boundaries,
\begin{equation}
  x'(\xi, t') = X^{\prime\lef}(t') + \xi \bigl[X^{\prime\lef}(t') - X^{\prime\rig}(t') \bigr]
\end{equation}
Here, $X^{\prime\lef}$ and $X^{\prime\rig}$ are the positions left and right of
the phase in question. We further use the shortcut $\Delta X' := X^{\prime\rig}
- X^{\prime\lef}$. The coordinate transformation adds a second advection\hyp{}like
term with velocity~$\partial_t x'(\xi,t')$ to the evolution equation, such that
we obtain for the model equations in each phase,
\begin{gather}
  \label{dxv_xi}
  \partial_\xi v^{0\prime}(\xi, t') = - \Delta X' q'(\phi), \\
  \label{dtphi_xi}
  \partial_{t'}\phi(\xi, t')
    = \frac{\partial_t x'(\xi,t')}{\Delta X'} \partial_\xi\phi
    - \partial_\xi \biggl[
        \frac{v^{0\prime}\phi}{\Delta X'}
      - \frac{D'(\phi)}{\Pe\:(\Delta X')^2} \partial_\xi\phi\biggr].
\end{gather}
Equations \eqref{dtX_num}, \eqref{dxv_xi}, and \eqref{dtphi_xi} are then solved
using the method of lines, with a finite\hyp{}volume discretization and linear
interpolation of~$\phi$ in space. The time\hyp{}stepping is done by a standard
explicit fourth\hyp{}oder Runge--Kutta algorithm~\cite{NumRecipes}.


\section{Prototype solutions: a single phase}
\label{sec:proto_single}
%
\begin{figure}[!b]%
  \centering
  \includegraphics{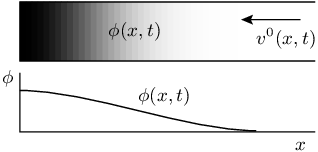}%
  \caption{Sketch of the typical single\hyp{}phase situation described in
  Sec.~\ref{sec:proto_single}}.%
  \label{fig:single_phase}%
\end{figure}%
In this and in the following sections we discuss several different solutions of
the above model equations, which are~\eqref{dxv_int} and \eqref{dtphi_int},
completed by the coefficient functions~$q(\phi)$ and $D(\phi)$ from
Eqs.~\eqref{diffcoeff} and \eqref{qmu}---Or, in their dimensionless versions,
Eqs.~\eqref{dxv_xi}, \eqref{dtphi_xi}, \eqref{diffcoeff_num}, and
\eqref{qmu_num}. We here choose two functions $q(\phi)$ and $D(\phi)$ and solve
the equations numerically, accompanied by analytical approximations. The aim is
to display typical solutions and to explore the variety of solutions by using
different functions $q(\phi)$ and $D(\phi)$. These solutions are meant to guide
future experiments and their interpretation, where the final target is to employ
the model for \emph{extracting} the functions $q(\phi)$ and $D(\phi)$ from
experimental data and to invert them to obtain the chemical potentials and the
phenomenological coefficients. This will then require to solve the inverse
problem of what we do in the present paper. We start with the description of
three typical situations with a single phase and smooth concentration profiles,
see also Fig.~\ref{fig:single_phase} for a sketch.
\subsection{Solvent only}
For setups containing only water and no solutes, the water loss is the
constant~$q(0)$, which has been measured experimentally by the total consumption
of water per time~\cite{LenJoaAjd07}. A situation with water only therefore
permits to identify the constants in expression~\eqref{qmu} for the
pervaporation coefficient. A~corresponding series of experiments with varying
parameters can be used to verify that the assumptions we made for~\eqref{qmu}
are indeed fulfilled. Possible variations can either be geometrical such as
changing the length~$L$ or the height~$h$ of the channel, or the membrane
thickness~$e$. Other variations concern the driving parameters, such as the
solute concentration~$\phi_L$ in the reservoir, or the principal control
parameter during the experiment, which is the external chemical
potential~$\chpot_w^\air$ of water in the air flow. This chemical potential is
controlled by the humidity of the air circulating through the top channel.
Equation~\eqref{qmu} predicts a linear dependence of the pervaporation on the
specific humidity~$c_w^\air$, with an offset,
\begin{equation}
  q(0) = \frac{1}{\vol_w} \frac{\barD}{eh} \biggl[
     \exp\Bigl\{\frac{m_w}{kT}\chpot_w(0)\Bigr\} -
     c_w^\air \exp\Bigl\{\frac{m_w}{kT}\chpot_w^{\air\ast}\Bigr\}
  \biggr],
\end{equation}
where the slope of the line and the offset may in principle depend on the
pressure. Quantitative information on this dependence would thus offer a
validation of the current model and yield insight into the thermodynamic
mechanisms inside the PDMS layer. Previous studies~\cite{RanDoy05} assumed that
the pervaporation is driven by a gradient of concentration and not of pressure,
as we did it in Sec.~\ref{sec:pervaporation}.
\subsection{Dilute solutions}
\label{sec:dilute}
\begin{figure}[!bt]%
  \centering
  \includegraphics{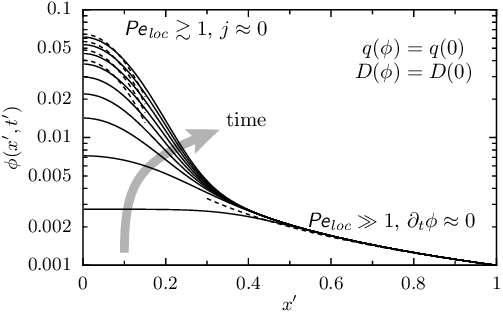}%
  \caption{Growth of a dilute solute density according to the two regimes
  ``Gaussian'' and ``hyperbolic ramp.'' The plot shows the numerical solution
  with parameters ($q'=1$, $D'=1$, $\Pe=100$, solid curves) together with the
  analytical approximation from Eqs.~\eqref{diluteapp:ramp} and
  \eqref{diluteapp:gauss} (dashed curves), see also
  Ref.~\cite{LenLonTabJoaAjd06}.}%
  \label{fig:dilute}%
\end{figure}%

The situation with dilute solutions, in which both the evaporation
coefficient~$q$ and the diffusion coefficient~$D$ are essentially constant, has
already been explored in Ref.~\cite{LenLonTabJoaAjd06}. Two different
regimes were identified in the solution, first a \emph{hyperbolic ramp}
corresponding to a steady uniform current at high Peclet numbers, and second a
Gaussian growing linearly in time, corresponding to the accumulation zone at
Peclet number equal one. Both regimes can be identified in Fig.~\ref{fig:dilute}
where a numerical solution of equations~\eqref{dxv_int}
and~\eqref{dtphi_int} is depicted together with the analytical approximation
from Ref.~\cite{LenLonTabJoaAjd06}. These approximations are given by the
solution of
\begin{equation}
  \label{diluteapp:ramp}
  \begin{gathered}[c]
  j'(1) = j'(x') \approx v^{0\prime}\phi = -q'x'\phi \\
  \Longrightarrow \phi(x') \approx -\frac{j'(1)}{q'x'}
  \end{gathered}
\end{equation}
for the hyperbolic ramp, and by the solution of
\begin{equation}
  \label{diluteapp:gauss}
  0 = j'(x') = -q'x'\phi - \frac{D'}{\Pe}\partial_{x'}\phi
\end{equation}
for the Gaussian profile. The width of the Gaussian hump thus scales as
$L/\sqrt{\Pe}$. The values given in Ref.~\cite{LenLonTabJoaAjd06} indicate a
Peclet number of~$100$.

\subsection{From dilute to dense single-phase solutions}
\label{sec:dense}

The mere model equations~\eqref{dxv_int} and \eqref{dtphi_int} include no
mechanism to prevent unphysical values $\phi(x, t) > 1$. We thus have to ask
what mechanisms the model provides to ensure $\phi\leq1$ in the dense limit when
$\phi$~approaches unity. For dense solutions, both coefficients $q(\phi)$~and
$D(\phi)$ exhibit non\hyp{}trivial dependences on~$\phi$ which provide two
independent physical mechanisms to keep $\phi$ below unity. These mechanisms are
described in the following paragraphs. Numerical examples for both are depicted
in Fig.~\ref{fig:bounded}.

\subsubsection{Vanishing pervaporation}

One mechanism for keeping the volume fraction bounded is governed by the
pervaporation coefficient~$q(\phi)$. Equation~\eqref{qmu} for this coefficient
implies an equilibrium value~$\phi^\eq$ at which the pervaporation vanishes. If
the concentration exceeds this value, the pervaporation direction is inverted,
which stabilizes $\phi$ around the equilibrium value, see Fig.~\ref{fig:mu}. In
combination with the boundary condition~\eqref{bc:v} for the velocity field we
find a \emph{flattening} of the concentration profile at~$\phi=\phi^\eq$, at
least in an interval starting with $x=0$. In this interval, the velocity remains
approximately zero. A numerical example of such a profile is given in
Fig.~\ref{fig:bounded}b.

\subsubsection{Diverging diffusion coefficient}

The other mechanism leading to volume fractions~$\phi(x,t)$ which strictly
cannot exceed unity is that the chemical potential of water diverges at
$\phi\to1$. In this case, the solvent becomes dilute in the solute, leading to
the standard logarithmic form of the chemical potential of water in the solute
concentration (cf.~Eq.~\eqref{chempoteq} or Eq.~\eqref{barchpotw_dilute}).
Together with the chemical potential also its derivative diverges in the limit
$\phi\to1$. In other words, the osmotic compressibility of the solute phase
diverges as well: The more the solute is concentrated, the more difficult it
becomes to concentrate it even further---and it cannot be concentrated beyond a
value $\phi^\td$ which is given by thermodynamics as the point where the
chemical potential diverges. $\phi^\td$~will be unity in most cases.

\begin{figure}[!bt]%
  \centering
  \includegraphics{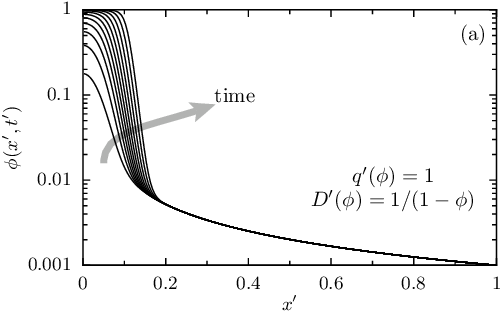}\\
  \includegraphics{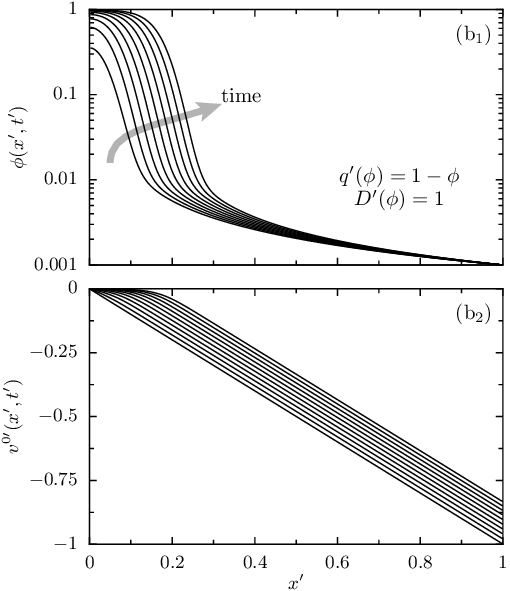}%
  \caption{Two different mechanisms to keep the density profile bounded from
  above: Panel~(a) shows the influence of a diverging diffusion coefficient
  $D'(\phi) = 1 / (1-\phi)$ while the pervaporation coefficient is constant
  ($q'=1$, $\Pe=10^3$). In this case, the velocity profile is linear for all
  times (not shown). In the two lower panels~(b) the diffusion coefficient is
  constant and the pervaporation coefficient varies with $\phi$ as $q'(\phi) =
  1-\phi$ ($D'=1$, $\Pe=10^3$). This leads to a time\hyp{}dependent velocity
  profile.}%
  \label{fig:bounded}%
\end{figure}

Mathematically speaking, the diffusive currents tend to flatten the spatial
concentration profile. In the limit of an infinite diffusion coefficient---which
is proportional to the derivative of the chemical potentials---the volume
fraction~$\phi(x,t)$ become arbitrarily flat. It will then not grow beyond the
value $\phi^\td$ at which the diffusion coefficient diverges. A~numerical
example for such a situation is depicted in Fig.~\ref{fig:bounded}a. This
flattening poses a bounding mechanism which is absent if the diffusion
coefficient stays finite or even vanishes.

It depends on the values of $\phi^\eq$ and $\phi^\td$ which of the two
mechanisms takes place first in a given mixture. In the present description,
however, which is based on thermodynamics, the logarithmic divergence must be
located at $\phi^\td=1$ which is always larger than the equilibrium
value~$\phi^\eq$. In a mixture of small solutes, which are of molecular size,
the value $\phi^\td=1$ should indeed be approachable. It is not approachable in
a system of rigid spheres in a much smaller solvent, where there is a
close\hyp{}packing limit $\phi^\td < 1$. In such systems with large solute
particles, the solvent can still pass through the holes between the spheres, and
the pervaporation never ceases~\cite{footnote2}. We acknowledge that such a
system cannot be described by our model---a deficiency which stems from the
treatment of the pressure, and which is therefore shared by any other
thermodynamic model that assumes mechanical equilibrium in form of the pressure
gradient balancing all external forces. In the following, we will therefore
focus the discussion on the case where the pervaporation vanishes before the
diffusion coefficient diverges,
\begin{equation}
  \phi^\eq < \phi^\td.
\end{equation}
For our treatment of $q$ and $D$, this implies that the divergence of $D$ is
not reached, because $q$ vanishes first. We thus expect the effect of
\emph{drying out} at least for very high densities. (Strictly speaking, this is
true only for non\hyp{}diverging~$\PhCoeff(\phi)$.) In dense solutions, especially
at the end of the pervaporation channel, we thus expect a flat concentration
profile with a nearly vanishing pervaporation and a large but finite diffusion
coefficient.

\begin{figure}[bt]
  \centering
  \includegraphics{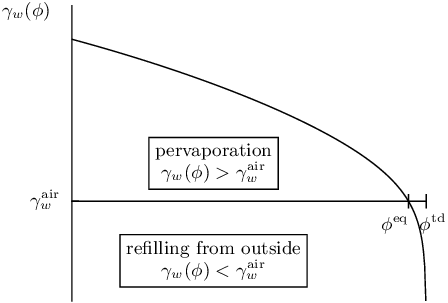}%
  \caption{A sketch of the chemical potential $\chpot_w(\phi)$ of water with the
  maximal concentration permitted by thermodynamics, $\phi^\td$. The
  pervaporation vanishes at the equilibrium value $\phi^\eq$ because of the
  equilibrium $\chpot_w(\phi^\eq) = \chpot_w^\air$.}%
  \label{fig:mu}%
  \bigskip
  \includegraphics{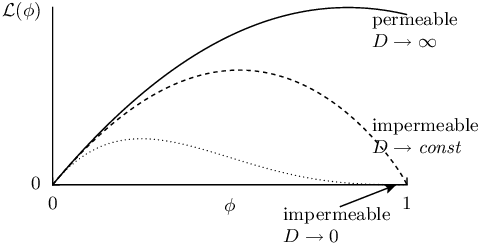}%
  \caption{Three possible functions~$\PhCoeff(\phi)$ for the phenomenological
  coefficient of inter\hyp{}diffusion: They lead to qualitatively different behavior
  of the diffusion coefficient at $\phi\to 1$.}%
  \label{fig:L}%
\end{figure}

We have to add a technical point here, as the diffusion coefficient depends not
only on the osmotic compressibility but also on the phenomenological
coefficient~$\PhCoeff(\phi)$. There are three different possibilities, which are
all sketched in Fig.~\ref{fig:L}: If~$\PhCoeff(1)$ is non\hyp{}zero, the diffusion
coefficient diverges as $\phi\to1$. This is the case which we have discusses so
far. In the cases with vanishing phenomenological coefficient~$\PhCoeff(1)=0$
the diffusion coefficient remains finite or eventually vanishes.%
%


\section{Prototype solutions: several phases}
\label{sec:proto_several}
\begin{figure}[!b]%
  \centering
  \includegraphics{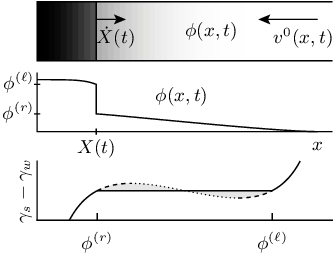}%
  \caption{Sketch of the typical situation with one moving phase boundary, described in
  Sec.~\ref{num:moving}}.%
  \label{fig:double_phase}%
\end{figure}

After the treatment of smooth concentration profiles such as the one in
Fig.~\ref{fig:single_phase}, we now continue with situations comprising several
distinct thermodynamic phases, see Figs.~\ref{fig:double_phase} and
\ref{fig:multi_phase}. As more and more solute is transported from the
reservoir into the channel the concentration rises and may lead to a phase
transition. The position of such a phase interface and its movement in time is
a convenient observable in experiments. The velocity of such an interface is
governed both by the advective and the diffusive currents around it, and it
thus presents a fingerprint of the mixture properties which have here been
reduced to the two coefficients $D(\phi)$ and $q(\phi)$. It does not provide as
much information on these coefficients as for example the full concentration
profile. The interesting question is how much information on the coefficients
can be extracted from the velocity of an interface as a function of time.

In addition to the model Eqs.~\eqref{dxv_int} and \eqref{dtphi_int} we now
take the positions $X_i(t)$ of the interfaces into account as unknown variables.
This type of problem is known as a \emph{free\hyp{}surface} or \emph{Stefan~problem}.
The interface movement is governed by Eq.~\eqref{int:dotX}, or in its dimensionless
variant by Eq.~\eqref{dtX_num}. The typical situation with one moving interface
is sketched in Fig.~\ref{fig:double_phase}.
\subsection{Two phases: Moving phase interfaces}
\label{num:moving}

To start with, we repeat the argument in Ref.~\cite{LenJoaAjd07} on the
experiment of solution of AOT~in water. It led to an exponential decrease of the
interface velocity, or equivalently to an interface position of the form
\begin{equation}
  \label{move:simple}
  \frac{X(t)}{L} = 1 - \exp\Bigl\{ -\frac{\phi_L \vol_w q(0)}{\Delta\phi} (t -
  t_0) \Bigr\}
\end{equation}
with an effective nucleation time~$t_0$. This result is based on several
assumptions: first, that all incoming solute current is transferred into the
interface movement. This means that the total solute current~$j(x,t)$ is a
step\hyp{}function, being zero in the dense phase (left\hyp{}hand side) and a non\hyp{}zero
constant in the dilute phase (right\hyp{}hand side). Equation~\eqref{move:simple}
assumes further that the velocity profile vanishes everywhere left of the
interface and that it is linear otherwise, $v^\vavg(x) = -\vol_w q(0) \bigl(x -
X(t)\bigr)$. With these two assumptions, the solute current which enters the
channel at the reservoir ($x=L$) is
\begin{equation}
  j(L,t) = \phi_L v^\vavg(L,t)
  = -\vol_w q(0) \phi_L [L - X(t)].
\end{equation}
The amount of solute coming from the reservoir thus depends on the position of
the interface. As all solute current arrives at the phase interface, we find its
movement from Eq.~\eqref{dtX_num} to be
\begin{equation}
  \dot X(t) = -\frac{j(X(t), t)}{\Delta \phi} = \frac{\vol_w q(0) \phi_L}{\Delta\phi} (L - X(t)).
\end{equation}
This differential equation leads to the solution~\eqref{move:simple}.

The two assumptions are quite strong, since the temporal change of the volume
fraction~$\phi$ does lead to a spatial change of solute current. Nevertheless,
the solution~\eqref{move:simple} has been successfully fitted to experimental
data, at least for short times. For longer times, however, it over\hyp{}estimated
the velocity. It is this deviation which we like to explore in the following.

For a more detailed analysis of the interface movement we have to develop a
global view on the spatial solute profile. In general, whatever value the solute
current has at the entrance of the channel (i.\,e.~the concentration~$\phi_L$
times the velocity there), it drops to zero at the end of the channel. We may
therefore identify three major possible growth mechanisms due to the spatial
change of the solute current along the channel: One possibility is the case we
discussed around Eq.~\eqref{move:simple}, where the solute current is constant
in both phases with a discontinuity at the interface. The current leads to the
growth of the dense phase and to the movement of the interface. The two other
possibilities are losses in the bulk of the dilute and the dense phases, leading
to a temporal change of the concentration profiles there.

Independent of the total solute current, the mixture velocity~$v^0$ may be zero
or not in the dense and dilute phases. In the following three subsections we
discuss three different combinations of vanishing\discretionary{/}{}{/}non\hyp{}vanishing
currents\discretionary{/}{}{/}velocities left and right of the interface.
\subsubsection{Solute accumulation at the phase boundary: the spatial profile in the dilute phase}
\label{sec:move:comoving}

We first investigate the spatial concentration profile which results from the
assumptions taken above for Eq.~\eqref{move:simple}, namely that all incoming
solute is accumulated at the phase interface. The dense phase plays no role,
i.\,e.~$v^\vavg(x) = 0$, $j(x) = 0$ for $x<X(t)$. The velocity of the interface
then depends only on the concentration profile in the dilute phase,
\begin{equation}
  \dot X(t) = \frac{D \partial_x \phi}{\Delta \phi}.
\end{equation}
In a system of coordinates moving together with the interface the evolution
equation for the density reads (compare with Eq.~\eqref{dtphi_xi}),
\begin{equation}
  \partial_t \phi = \dot X \partial_x \phi - \partial_x j.
\end{equation}
A steady solution in this moving frame corresponds to a uniform current $j_0 =
-\Delta\phi \dot X$ in the reference frame. The uniformity of the current
corresponds to the above mentioned assumption that all solute entering the
dilute phase leaves it at the interface. We then find the profile stationary
with respect to the moving frame as the solution of
\begin{equation}
  (v^\vavg - \dot X) \phi - D\partial_x\phi = j_0
\end{equation}
As this equation is to be solved in the dilute phase, we may assume constant
coefficients $q(\phi) \approx q(0)$ and $D(\phi) \approx D(0)$ which allows to
find the spatial profile
\begin{multline}
  \label{move:stat}
  \phi(X(t) + y) = \exp\Bigl\{-\Bigl(\frac{y+\beta}{\alpha}\Bigr)^2\Bigr\}
  \times \\
  \biggl[
    \phi^\rig \exp\Bigl\{\frac{\beta^2}{\alpha^2}\Bigr\}  +
    2 \Delta\phi\frac{\beta}{\alpha}
    \Bigl[
    G\Bigl(\frac{y+\beta}{\alpha}\Bigr) -
    G\Bigl(\frac{\beta}{\alpha}\Bigr)
    \Bigr]
  \biggr],
\end{multline}
with the shortcuts
\begin{equation}
  \label{comoving:scaling}
  \alpha := \sqrt{\frac{2D}{\vol_w q}} = L\,\sqrt{\frac{2}{\Pe}},\quad
  \beta := \frac{\dot X}{\vol_w q},\quad
  G(x) := \int_0^x e^{\xi^2} d\xi.
\end{equation}
\begin{figure}[!bt]%
  \centering
  \includegraphics{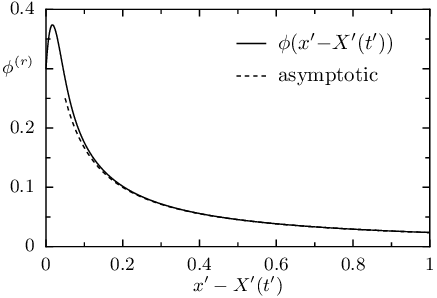}
  \caption{Visualization of the solution \eqref{move:stat} which is stationary in the
  frame moving with the interface. The parameters are arbitrary ($\Pe=10^3$,
  $\dot X'=0.05$, $\phi^\rig=0.3$, $\phi^\lef=0.8$).}%
  \label{fig:statsol}%
\end{figure}%
An example of this solution is shown in Fig.~\ref{fig:statsol}. It exhibits the
same two regimes as the dilute equation described above in Sec.~\ref{sec:dilute}
and in Fig.~\ref{fig:dilute}: There is a hump dominated by diffusion and a
hyperbolic ramp.

The solution from Eq.~\eqref{move:stat}, depicted in Fig.~\ref{fig:statsol}
gives insight into the implications of the taken assumptions: It shows that the
diffusive hump \emph{exceeds} the binodal concentration value~$\phi^\rig$.
Evidently, there is a zone of higher concentration which moves together with
the phase interface, and which continuously feeds the dense phase. As long as
both the velocity and the total solute current vanish at the interface, the
diffusive hump \emph{must} exceed the value~$\phi^\rig$ since there must be a
negative current causing the interface to move to the right. The only
contribution to the solute current is the one caused by the concentration
gradient, thus having to be positive. The assumptions $j(X,t)=0$ and
$v^\vavg(X,t)=0$ create a situation in which it is in principle possible that
the peak which is clearly visible in Fig.~\ref{fig:statsol} reaches a spinodal
value, which is larger than $\phi^\rig$ but smaller than $\phi^\lef$. As the
concentration reaches the spinodal, it will undergo local phase changes and
exhibit local blobs of a metastable phase. Such blobs have occasionally been
observed in the experiments using a KCl~solution~\cite{LenLonTabJoaAjd06}.

The diffusive hump in Fig.~\ref{fig:statsol} appears pronounced due to the large
Peclet number chosen in this example. The Peclet number renders the length
scale~$\alpha$ small, which reduces the lateral extension of the hump and
increases its maximum value. In experimental realizations, its height is limited
by several factors, first by the overall Peclet number, second by the spinodal
concentrations which enforce a local phase transition instead of the smooth
maximum, and third by the pervaporation in the dense phase. For a non\hyp{}vanishing
velocity $v^\vavg(X,t)$ the necessity of having a positive slope
$\partial_x\phi$ at $x=X(t)$ gradually vanishes, allowing the height of the
maximum to decrease or to vanish completely.

We continue to discuss the solution~\eqref{move:stat} by showing how the
result~\eqref{move:simple} follows from the spatial form of the solute
concentration. The hyperbolic ramp, which is the asymptotic solution for
large~$x$, here reads (in rescaled units)
\begin{equation}
  \label{move:asymp}
  \phi(X(t) + y) \sim \frac{\Delta\phi}{1 + y'/\dot X'}.
\end{equation}
We may use the asymptotic form to calculate an approximate velocity of the
interface: Of course, this result corresponds to a stationary solution in a
reference frame moving at constant velocity. Therefore, the input concentration
at the end of the channel must be increased consistently in order to find this
solution. Reversely, if the input concentration is held constant at $\phi(L) =
\phi_L$, the current arriving at the phase interface decreases in time. As long
as the diffusive hump does not reach the inlet of the channel, this change of
current is small and we may use $\phi(L)$ as a slowly varying parameter. Setting
$\phi(L) = \phi_L$ in the asymptotic solution~\eqref{move:asymp} then yields
essentially the same expression for the evolution of the interface position as
in equation~\eqref{move:simple},
\begin{equation}
  \label{move:asymptotic}
  X'(t) = 1 - \exp\Bigl\{-\frac{\phi_L}{\Delta\phi - \phi_L} (t'-t'_0)\Bigr\}.
\end{equation}
%
\subsubsection{Solute accumulation at the phase boundary: the velocity field in the dense phase}

As a next approximation we relax the assumption of a vanishing velocity left of
the interface. However, we continue to keep the total current zero in the dense
phase. Also the dilute phase is passive, such that all incoming solute is
accumulated at the interface.

As discussed above in section~\ref{sec:dense} the velocity and density profiles
in dense phases can behave in two qualitatively different ways: One corresponds
to the vanishing of~$q(\phi)$ and the other to the divergence of~$D(\phi)$. We
make the same distinction here. The first case, with vanishing~$q(\phi)$, is
characterized by a flat velocity profile near the end of the channel. It is
likely to lead to a density profile which has maximum density from the end of
the channel up to the interface, where it reaches a small region with
non\hyp{}vanishing velocity. After a transient time, the density profile in a small
region could become stationary in a reference frame moving together with the
interface. This solution supports a mixture velocity at the interface that is
\emph{constant} in time. The position of the interface is then given by
\begin{equation}
  \label{move:constv}
  \frac{X(t)}{L} = \Bigl(1 - \frac{v(X)}{L\vol_w q(0)}\Bigr)
    \Bigl(1 - \exp\Bigl\{ -\frac{\phi_L \vol_w q(0)}{\Delta\phi} (t-t_0)
    \Bigr\}\Bigr).
\end{equation}
The dilute phase is here characterized by a constant $q$ and $D$.

In the second case, in which the diverging diffusion coefficient leads to a
bounded density profile, the velocity field needs not to vanish except at the
very end of the channel. For simplicity, we adopt a constant non\hyp{}zero value
for the pervaporation in the dense phase, $q(\phi) \approx q(1) < q(0)$. This
solution leads to a piecewise linear mixture velocity and to the interface
position
\begin{equation}
  \label{move:linearv}
  \frac{X(t)}{L} = \frac{q(0)}{q(0) - q(1)}
    \Bigl(1 - \exp\Bigl\{ -\frac{\phi_L \vol_w q(0)}{\Delta\phi} (t-t_0)
    \Bigr\}\Bigr).
\end{equation}

None of the solutions \eqref{move:constv}~and \eqref{move:linearv} can explain
the velocity decrease found in the AOT~experiments. Instead, both velocities are
larger than the initial proposition solution~\eqref{move:simple}. Note that we
still kept the assumption $j(x)=0$ in the dense phase, which corresponds to a
large diffusion coefficient. It must therefore be concluded that in the case of
AOT either the diffusion coefficient is not sufficiently large, or the velocity
decrease stems from the dilute phase.
\subsubsection{Solute accumulation also in the dilute phase}
\label{sec:move:num}

In both preceding subsections we have kept the assumption that all incoming
solute contributes to the movement of the interface. We now continue with the
more general case where there may be a temporal change of the concentration
profile such that there is a gradient of solute current in the dilute phase. The
dense phase is still assumed to be passive (no pervaporation).
\begin{figure}[!bt]%
  \centering
  \includegraphics{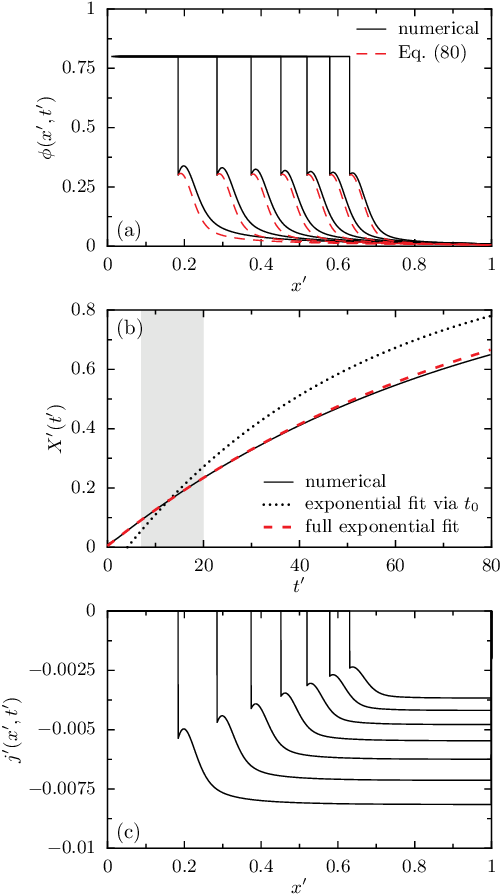}%
  \caption{Numerically determined density profile~$\phi(x', t')$ containing a
  single moving phase interface. The binodal density values at the interface are
  $\phi^\lef=0.8$ and $\phi^\rig=0.3$; at the channel entry $\phi_L=10^{-2}$;
  $\Pe=10^3$. Panel~(a) shows the density profile, panel~(b) the position of the
  interface as a function of time, and panel~(c) the evolution of the spatial
  current profile.}%
  \label{fig:moving}%
\end{figure}%

This more general situation cannot be described analytically but has to be
treated numerically. A numerical test of the situation with a dense phase
growing into a dilute one exhibits the limitations of the assumption taken
above in Sec.~\ref{sec:move:comoving}. Figure~\ref{fig:moving} shows the full
numerical solution of the model. The velocity in the dense part has been kept
zero ($q=0$), leading to a flat density profile there. The dilute part of the
solution agrees qualitatively with the graph given in Fig.~\ref{fig:statsol}.
There are, however, deviations. The plot shows the solutions of
Eq.~\eqref{move:stat} with the true position and the true velocity of the
interface taken from the numerical solution. Panel~b quantifies the difference
between the approximation from Sec.~\ref{sec:move:comoving} and the true
solution. It shows the position of the interface as a function of time. A fit
with the exponential function~\eqref{move:simple} with the known value for
$\phi_L\vol_w q/\Delta\phi$ and $t_0$ as the fitting parameter proves that the
approximation does not work in this example. A second fit for both parameters
indicates that the resulting function is still exponential, but with a
different timescale than assumed in the approximation~\eqref{move:simple}. In
both fittings only the data in the gray\hyp{}shaded area has been used. The fit
quantifies that around $60\%$~of the incoming solute is accumulated at the
interface and leads to the growth of the dense phase. The rest remains in the
dilute phase. Panel~\ref{fig:moving}c visualizes this ratio directly.
Evidently, the current has not the form of a step function, but takes on
different values at the interface and at the channel entry. This clearly
falsifies the assumption of a uniform current in the dilute phase.

Upon varying the parameters of the example, we found that the concentration
$\phi^\rig$ right of the interface has the largest influence on the deposition
of current. For $\phi^\rig=0.05$ nearly all of the incoming current was found to
accumulate at the phase boundary---despite a well developed diffusive hump
similar to the one in Fig.~\ref{fig:moving}a, and despite the fact that the
solute current still exhibited a pronounced hump. The current apparently
overflows the diffusive hump completely, if $\phi(x,t)$ has already reached the
value $\phi^\rig$ before the hump starts. We may quantify this by stating that
approximation~\eqref{move:simple} is valid if
\begin{equation}
  \phi^\rig \leq \sqrt{\Pe}\: \frac{L-X}{L}\: \phi_L.
\end{equation}
Here, we assumed the width of the diffusive hump to scale as $L/\sqrt{\Pe\,}$,
see the definition of $\alpha$~in \eqref{comoving:scaling}, and also
Ref.~\cite{LenLonTabJoaAjd06}.
\subsection{Several phases: Phase slab thicknesses in stationary solutions}
\label{sec:severalphases}
\begin{figure}[!b]%
  \centering
  \includegraphics{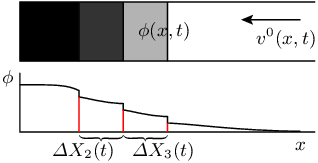}%
  \caption{Sketch of the situation with several phase slabs exhibiting
  characteristic extents, as described in
  Sec.~\ref{sec:severalphases}}.%
  \label{fig:multi_phase}%
\end{figure}%

We now return to the introductory example of Sec.~\ref{sec:intuitive} with four
phases, one of them dilute. This situation is sketched in
Fig.~\ref{fig:multi_phase}. The aim here is to present an example how the
thicknesses of phase slabs are governed by the thermodynamic and dynamic
properties of all the phases.

For the solution of AOT~in water, a remarkable observation has been reported in
Ref.~\cite{LenJoaAjd07}: First, the same phases as in equilibrium have been
found in the evaporation channel. They were identified by their polarizability
and birefringence, showing hexagonal, cubic and lamellar internal structures for
the dense phases, while the dilute phase was isotropic. The spatial pattern
showed that the two enclosed slab thicknesses (of the cubic and the lamellar
phases) were of comparable extension. This result is remarkable because these
two phases occupy a very different range of $\phi$-values in the equilibrium
phase diagram. There, the lamellar phase appears much broader than the cubic
phase, see the supplementary material of Ref.~\cite{LenJoaAjd07}. We conclude that
there is apparently a much larger concentration gradient in the lamellar phase
than in the cubic phase. This larger concentration gradient was stable in the
experiment, even when all dense phases were continuously fed with more incoming
solute. After a short transient time, the interfaces bounding the cubic and the
lamellar phase moved together at constant distance.

We now show that our model possesses qualitatively the same solutions. We also
observe stable thicknesses of comparable size for the lamellar and the cubic
phases, if we choose the chemical potential and the phenomenological
coefficients well. The strategy here is to adopt simple model functions for
$\chpot_w(\phi)$ and $\PhCoeff(\phi)$. They are then transferred into diffusion
coefficient~$D(\phi)$ and pervaporation coefficient~$q(\phi)$ using
Eqs.~\eqref{qmu_num} and \eqref{diffcoeff_num} which are then used in the
stationary model equations.

We here adopt the most simple models for the chemical potential and for the
phenomenological coefficients. The functions used below are displayed in
Fig.~\ref{fig:sharp1}. The values of~$\phi$ are taken from the equilibrium phase
diagram of~AOT (see supplementary material of Ref.~\cite{LenJoaAjd07}). The
regions shaded in gray indicate phase coexistence, where the chemical potentials
are constant. Within the most dense phase we use a logarithmic behavior of the
form $\chpot^\prime_w(\phi) \propto \ln(1-\phi)$, whereas in the other three
phases we linearly connect some coexistence values which have been chosen as
model parameters. The phenomenological coefficient has been chosen constant in
each phase except in the dilute phase where a linear function of~$\phi$ is
required (see Sec.~\ref{sec:permeab}). The coefficients $q'(\phi)$ and
$D'(\phi)$ follow from equations \eqref{qmu_num} and \eqref{diffcoeff}, see
Fig.~\ref{fig:sharp2}.
\begin{figure}[!tb]%
  \centering
  \includegraphics{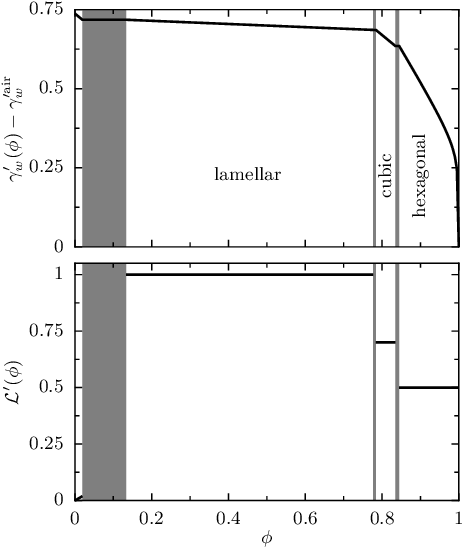}%
  \caption{Simple model functions for the rescaled chemical potential difference
  of water across the PDMS layer and for the phenomenological coefficient, both
  as functions of the volume fraction~$\phi$. Note the different slopes of the
  chemical potential in the lamellar and the cubic phases which lead to the same
  slab thickness in Fig.~\ref{fig:sharp3}.}%
  \label{fig:sharp1}%
  \centering
  \includegraphics{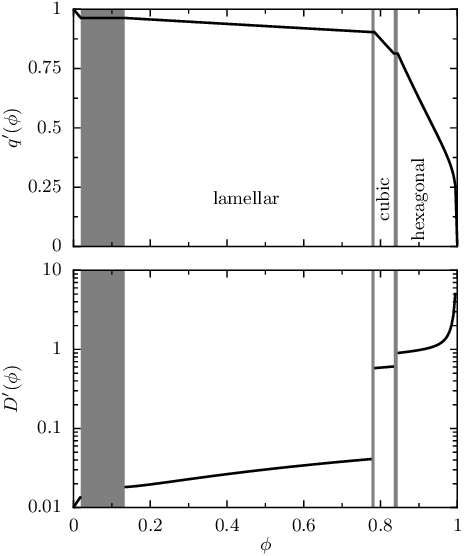}%
  \caption{The rescaled coefficients for evaporation~$q'(\phi)$ and for
  diffusion~$D'(\phi)$ resulting from the model in Fig.~\ref{fig:sharp2}.}%
  \label{fig:sharp2}%
\end{figure}%
\begin{figure}[!tb]%
  \centering
  \includegraphics{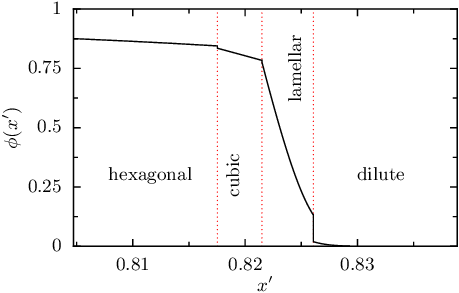}%
  \caption{An example solution of the stationary equations in which the
  thickness of phase slabs is controlled by the slope of chemical potential in
  Fig.~\ref{fig:sharp1}. Note the equal thickness of the lamellar and the cubic
  phases.}%
  \label{fig:sharp3}%
  \centering
  \includegraphics{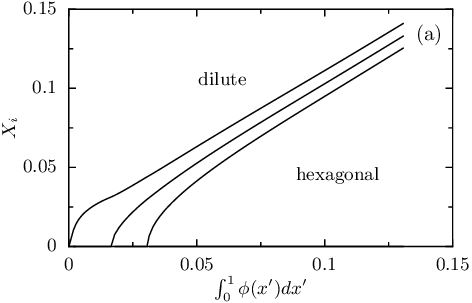}
  \includegraphics{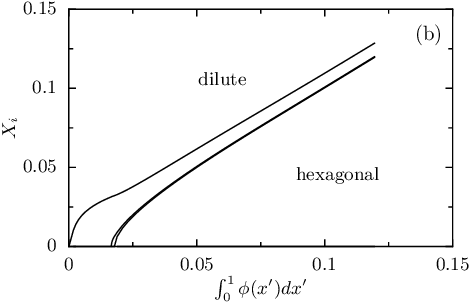}%
  \caption{The development of phase thicknesses for (a)~the model given in
  Fig.~\ref{fig:sharp1}; and in (b)~for a chemical potential in the cubic phase
  having the same slope as the one in the lamellar phase.}%
  \label{fig:sharp4}%
\end{figure}%

Figure~\ref{fig:sharp3} shows a typical profile of the volume fraction of
solute, $\phi(x')$, as it results from the model functions given in
Figs.~\ref{fig:sharp1} and \ref{fig:sharp2}. The lamellar and cubic phases are
indeed found to have approximately the same extent. It is mainly the different
slope of the chemical potential in Fig.~\ref{fig:sharp1}a which has been
adjusted manually to achieve the comparable extent of the two phases. The
physical content of these slopes are the very different osmotic
compressibilities of these two phases. Of course, also the phenomenological
coefficient plays a role. We do not claim that the presented chemical
potentials are the only ones which may lead to comparable phase extensions.
However, it is evident in this very example that the osmotic compressibilities
of the chemical potential directly influence the extension of the phase slabs.
The profile in Fig.~\ref{fig:sharp3} has been calculated as the stationary
solution of the model equations~\eqref{dxv_num} and \eqref{dtphi_num}. The use
of the stationary equations will be explained in the following.

\subsubsection{Stationary and quasi-stationary solutions}

The full evolution of the phase thicknesses in time is subject to many
parameters including not only the model parameters for $D(\phi)$ and $q(\phi)$
and the incoming volume fraction and the Peclet number. It depends also on the
initial state, or in other words, on the precise conditions of phase creation at
the end of the channel. We do not want to treat the issues of phase creation
here, the more as a reasonably fast numerical method for this parameter regime
is not at hand. Instead, we employ a quasi\hyp{}stationary treatment, which is
focused on the main features of the phase thicknesses. The idea is to assume a
very small input current and to assume at any time an equilibrated stationary
profile. Such a stationary profile is shown in Fig.~\ref{fig:sharp3} for the
model functions of Fig.~\ref{fig:sharp1}. Note the slab thicknesses of the
lamellar and the cubic phases which are of the same extent. This behavior has
been achieved by choosing the slopes of the chemical potential in
Fig.~\ref{fig:sharp1}, which is $20$~times larger in the cubic phase than in the
lamellar one. Of course, the variation of the chemical potential also has its
influence on the evaporation. We thus cannot clearly separate between thickness
due to a variation of diffusion and due to modified pervaporation. This
ambiguity is a systematic issue in the interpretation of the experimental
results, and it is reflected in our model by the two coefficients $D$ and~$q$.

In order to obtain the ensemble of all boundary positions, which is given in
Fig.~\ref{fig:sharp4}, we parametrized the solutions by the total amount of
solute $\int\phi(x')\,dx'$. In a quasi\hyp{}stationary setting with extremely small
input current, this amount may be understood as time multiplied by the input
current. A numerical test of the full time\hyp{}dependent equations revealed similar
lines as a function of time (not shown). It can be seen that the interior slab
thicknesses relax to constant values, and that only the last, the hexagonal
phase continues to grow. This behavior stems from the vanishing pervaporation in
the hexagonal phase which leads at the interfaces to a mixture velocity constant
in time. If the velocity does not vanish in the dense phase, then the interior
thicknesses do not truly become stationary.

In order to demonstrate the effect of the slopes of the chemical potential on
the slab thicknesses, Fig.~\ref{fig:sharp4}b presents the result for a modified
model chemical potential. We took the curve in Fig.~\ref{fig:sharp1}a and
changed the slope in the cubic phase to the slope in the lamellar phase. This
situation implies the same osmotic compressibilities in both phases, and also a
comparable spatial concentration gradient. As a result, the cubic phase has
shrunk and is hardly visible in Fig.~\ref{fig:sharp4}b.

The numerical solutions in Figs.~\ref{fig:sharp3} and \ref{fig:sharp4} have been
created by a Runge--Kutta algorithm solving the stationary
equations~\eqref{dxv_int} and~\eqref{dtphi_int}. On the analytical side, they
may be solved formally by integration over~$\phi$. The stationary equations can
be rewritten such that
\begin{gather}
  \frac{d}{dx}(v^\vavg)^2 = - \frac{D(\phi) \partial_x\phi}{\phi} \vol_w q(\phi)
  \quad\text{and}\\
  \frac{dx}{d\phi} = \frac{D(\phi)}{v^\vavg \phi}.
\end{gather}
The formal solution of the problem then reads
\begin{gather}
  v^\vavg(x, x_0) = -\biggl[{\textstyle \bigl(v^\vavg(x_0)\bigr)^2 + 2 \vol_w\int\limits_{\phi(x)}^{\phi(x_0)}
   \frac{D(\xi)q(\xi)}{\xi} d\xi}\biggr]^\frac{1}{2} \\
  x = x_0 + \int\limits_{\phi(x)}^{\phi(x_0)} \frac{D(\varphi)}{\varphi}
    \biggl[{\textstyle \bigl(v^\vavg(x_0)\bigr)^2 + 2 \vol_w\int\limits_{\varphi}^{\phi(x_0)}
       \frac{D(\xi)q(\xi)}{\xi} d\xi}\biggr]^{-\frac{1}{2}} d\varphi.
\end{gather}

This expression describes in principle any stationary solution, but in the
unusual way~$x(\phi)$. This formulation is especially convenient for the
determination of phase slab thicknesses. For the first phase at the end of the
evaporation channel, the expression simplifies due to the boundary condition
$v^\vavg(0) = 0$. In case of the phase interface $x=X(t)$ the lower bound of the
integrals is fixed to the binodal value~$\phi^\lef$ of the phase transition. We
then find
\begin{gather}
  v^\vavg(X(t)) = -\biggl[{2\vol_w \int_{\phi^\lef}^{\phi(0,t)}
   \frac{D(\xi)q(\xi)}{\xi} d\xi}\biggr]^\frac{1}{2} \\
  X(t) = \int_{\phi^\lef}^{\phi(0,t)} \frac{D(\varphi)}{\varphi}
    \biggl[{2\vol_w \int_{\varphi}^{\phi(0,t)}
       \frac{D(\xi)q(\xi)}{\xi} d\xi}\biggr]^{-\frac{1}{2}} d\varphi
\end{gather}

\subsubsection{Phase nucleation}

The treatment of the phase thickness in the present section resembles to some
extent the treatment of nucleation as it has been presented by Evans, Poon, and
Cates~\cite{EvaPooCat97}. Both include the assumption of local equilibrium,
leading to a diffusion\hyp{}type equation, while the total system is out of
equilibrium. Both allow the occurrence of metastable phases which are absent in
equilibrium. Here, these phases are the dense ones growing because of the
non\hyp{}zero solvent flow; in Ref.~\cite{EvaPooCat97} it is a freshly nucleated
metastable phase which grows because the whole system is quenched in
temperature.

We have not included the details of phase nucleation near the end of the
channel, as it would require a more advanced numerical treatment. The general
framework, however, should be applicable to the nucleation in a similar manner
as in Ref.~\cite{EvaPooCat97}.

\section{Summary}
\label{sec:summary}
We have derived a model for the behavior of binary solute--solvent mixtures in
recently developed microevaporator devices. The final target of the model is to
allow an interpretation of experimentally observed data in terms of
well\hyp{}defined thermodynamic and dynamic properties of the mixture which are
measurable also by different means. As a first step towards this goal we have
developed the model equations~\eqref{dxv_int} and \eqref{dtphi_int} which
relate the phase behavior in the microevaporator to an inter\hyp{}diffusion
coefficient~$D$ and another coefficient~$q$ quantifying the pervaporation of
solute. We have shown in Eqs.~\eqref{diffcoeff} and \eqref{qmu} how these two
coefficients depend on the chemical potentials in the mixture and on the
phenomenological coefficient for inter\hyp{}diffusion. This relation has then been
extended to make also the connection to other quantities such as the
coefficients of sedimentation, permeability and osmotic compressibility.

The presented model is focused on the transport equations of solute and solvent
mass in binary mixtures, using the hypothesis of local thermodynamic
equilibrium. We provide in detail all the necessary assumptions to reduce the
description to two variables, one of which is the volume fraction of solute, the
other one is the volume\hyp{}averaged mixture velocity. We further assume
incompressibility of the mixture and reduce the spatial dimensions to the one
parallel to the evaporation channel.

We solved the model equations in a number of typical limit cases which still
allow analytical approximations and compared these with full numerical
solutions. In particular we found good agreement in the case of dilute
solutions, where the diffusion and pervaporation coefficients may be treated as
constants. In dense binary mixtures, where this is no longer the case, we were
able to show that there are two qualitatively different mechanisms which both
lead to a saturation of the concentration at its maximal possible value. One of
them is due to vanishing evaporation at an equilibrium concentration. The other
case comes from the singular behavior of the chemical potential of water in this
limit.

We have further analyzed the growth velocity of a dense phase into a dilute one
and have found several expressions which allow to extract quite some
information on the properties of the mixture only from tracking a single
interface between two phases. Further, we have analyzed the resulting
thicknesses of phase slabs in the microevaporator in a stationary setting. In
particular, the experimentally observed case of equal phase thicknesses has
been regarded. We were able to explained this remarkable fact as the result of
very different osmotic compressibilities in the two phases.

In total, the results presented here show that the interpretation of the
experimental results is possible but far from trivial. In all cases, there is
some ambiguity in the interpretation of the observed behavior: As the physical
properties of the mixture are summarized in two coefficients, also the reasons
for the flattening at high solute concentrations are twofold. In a similar
manner, the thicknesses of the phase slabs are not only governed by their
thermodynamic properties, but also by the phenomenological coefficient
containing dynamic information.

It is precisely this twofold information contained in the microevaporator data
as well as in the two coefficients~$D$ and $q$ which presents the core of the
present work. We are convinced that the presented model establishes a good first
compromise between the complexity of the phenomenon being out of equilibrium and
the need for simplicity in the equations. Our model will help on the
experimental plan to identify sets of measurements which allow to lift the
ambiguity in the coefficients. It is the detailed derivation of the model and
the numerical treatment of the equations which allows to improve the
interpretation of the experimentally measured quantities and to extract from
them the chemical potentials and the transport coefficients as full functions of
the solute concentration.
\par\medskip\emph{Acknowledgements:} 
We are grateful to J.~Leng for valuable discussions on the experiments. For
comments on the manuscript we thank K.~Sekomoto and D.~Lacoste. This work was
financed by the German~\emph{DFG} via grant Schi/1-1 and by the
French~\emph{ANR} via project~``Scan~2''.



\begin{thebibliography}{99}

\bibitem{industry}
  W. Machtle, in: S. E. Harding, A. J. Rowe, J. C. Horton (Eds.),
  \textit{Analytical Ultracentrifugation in Biochemistry and Polymer Science}
  (The Royal Society of Chemistry, Cambridge 1992) 147--175.

\bibitem{cells}
  Q. G. Wang, H. D. Tolley, D. A. LeFebre, and M. L. Lee,
  Anal.\ Bioanal.\ Chem.~\textbf{373}, (2002) 125.

\bibitem{RusSavSch89}
  W.\,B.\ Russel, D.\,A.\ Saville, and W.\,R.\ Schowalter,
  \textit{Colloidal Dispersions}
  (Cambridge University Press, Cambridge, 1989).

\bibitem{PepEllWor05}
  S.~S.~L.~Peppin, J.~A.~W.~Elliot, and M.~G.~Worster,
  Phys.~Fluids \textbf{17}, (2005) 053301.

\bibitem{LebLewSch02}
  J. Lebowitz, M. S. Lewis, P. Schuck,
  Protein Sci.~\textbf{11}, (2002) 2067.

\bibitem{Paul04}
  D. R. Paul,
  J.~Membrane Sci.~\textbf{241}, (2004) 371.

\bibitem{LenLonTabJoaAjd06}
  J. Leng, B. Lonetti, P. Tabeling, M. Joanicot, and A. Ajdari,
  Phys. Rev. Lett \textbf{96}, (2006) 084503.

\bibitem{LenJoaAjd07}
  J. Leng, M. Joanicot, and A. Ajdari,
  Langmuir \textbf{23}, (2007) 2315.

\bibitem{SalLen08}
  J.-B.~Salmon, J.~Leng,
  to~appear.

\bibitem{Fraden}
  J.~Shim, G.~Cristobal, D.~R.~Link, T.~Thorsen, Y.~Jia, K.~Piattelli, and S.~Fraden,
  J.~Am.~Chem. Soc. \textbf{129}, (2007) 8825.

\bibitem{Baitz}
  B.~T.~C.~Lau, C.~A.~Baitz, X.~P.~Dong, and C.~L.~Hansen,
  J.~Am.~Chem. Soc. \textbf{129}, (2007) 454.

\bibitem{GroMaz84}
  S.\,R. de Groot and P. Mazur,
  \textit{Non-Equilibrium Thermodynamics}
  (Dover Publications, New York 1984).

\bibitem{JosHuaHu96}
  D. D. Joseph, A.~Huang, and H.~Hu,
  Physica~D \textbf{97}, (1996) 104.

\bibitem{footnote3}
  The comparison of the two terms in Eq.~\eqref{dtphi} requires
  several individual estimations: (A)~The pressure gradient is approximated by
  the one in a Poiseuille flow in the same rectangular channel. This should be a
  good approximation for a dilute solution at the entrance of the channel near
  the reservoir. We take typical values from Ref.~\cite{LenLonTabJoaAjd06},
  namely $v^0(L)=13\mu\text{m}/\text{s}$, $w=200\mu\text{m}$, $h=20\mu\text{m}$, together with the
  viscosity of water, $\eta=10^{-3}\text{kg}/(\text{m}\text{s})$. This leaves us with a
  pressure gradient of
  %
  \begin{equation}
    \frac{\Delta P}{L} = \eta\:\alpha\big(w/h\big)\:\frac{v^0}{wh} \approx 3.9
    \frac{\text{Pa}}{\text{m}}
  \end{equation}
  %
  with the shape-factor $\alpha(100)\approx 1200$ taken from
  Ref.~\cite{MorOkkBru05}. As a further approximation (B)~we take the chemical
  potentials in their dilute limit, where $\chpot_s$ becomes logarithmic in
  $c_s$, yielding
  %
  \begin{equation}
    \frac{\partial(\chpot_s-\chpot_w)}{\partial c_s} \approx
    \frac{kT}{m_s}\frac{1}{c_s}.
  \end{equation}
  %
  We also anticipate~(C) the ``hyperbolic ramp'' solution at the entrance of the
  channel (see Sec.~\ref{sec:dilute} and Ref.~\cite{LenLonTabJoaAjd06} for an
  experimental evidence), $c_s(x,t)\approx c_s(L) L / x$. This approximation
  leaves us with the concentration gradient of the order $|\grad c_s|\approx
  c_s(L)/L$. Putting all pieces together, the ratio of the two terms in
  Eq.~\eqref{dtphi} is
  %
  \begin{equation}
    \label{cit:ratio}
    \frac{(\vol_s{-}\vol_w) |\grad P|}{\frac{\partial(\chpot_s-\chpot_w)}{\partial
    c_s} |\grad c_s|}
    \approx (\vol_s{-}\vol_w) \frac{Lm_s}{kT}\:3.9\frac{\text{Pa}}{\text{m}}.
  \end{equation}
  %
  If we say that the solute is ten times lighter than the solvent water
  (approximation D), the difference of specific volumes will be dominated by
  that factor. Together with a molecular weight of a few hundred atom units, we
  end up with the factor in Eq.~\eqref{cit:ratio} to be less than $10^{-3}$, which
  justifies the approximation in Sec.~\ref{sec:pressgrad}.

\bibitem{NumRecipes}
  W.\,H.~Press and S. A. Teukolsky and W. T. Vetterling and B. P. Flannery,
  \textit{Numerical Recipes in~C}
  (Cambridge University Press, Cambridge 1992).

\bibitem{RanDoy05}
  G.~C.~Randall, P.~S.~Doyle,
  Proc. Natl. Acad. Sci. USA. \textbf{102}, (2005) 10813.

\bibitem{footnote2}
  The pervaporation may cease if the spheres exhibit some compressibility which
  enters the chemical potential.

\bibitem{EvaPooCat97}
  R. M. L. Evans, W. C. K. Poon and M. E. Cates,
  Europhys.\ Lett.~\textbf{38}, (1997) 595.

\bibitem{MorOkkBru05}
  N.~A.~Mortensen, F.~Okkels, and H.~Bruus,
  Phys.\ Review~E~\textbf{71}, (2005) 057301.

\end{thebibliography}
\end{document}